\documentclass[useAMS,usenatbib]{mn2e}
\usepackage{epsfig}
\usepackage{subfigure}
\usepackage{amssymb}
\usepackage{color}
\usepackage{booktabs}                  
\newcommand{\bn}{\begin{enumerate}}
\newcommand{\en}{\end{enumerate}}
\newcommand{\bi}{\begin{itemize}}
\newcommand{\ei}{\end{itemize}}

\def\gtorder{\mathrel{\raise.3ex\hbox{$>$}\mkern-14mu
    \lower0.6ex\hbox{$\sim$}}}
\def\ltorder{\mathrel{\raise.3ex\hbox{$<$}\mkern-14mu
    \lower0.6ex\hbox{$\sim$}}}

\title[Comparing AMR and SPH Approaches in Direct Collapse Modeling]
{Direct Collapse to Supermassive Black Hole Seeds: Comparing the AMR
and SPH Approaches}

\author[Yang Luo, Kentaro Nagamine, and Isaac Shlosman]
{Yang Luo$^{1}$\thanks{E-mail:yluo@vega.ess.sci.osaka-u.ac.jp},
Kentaro Nagamine$^{1,2}$\thanks{E-mail:kn@vega.ess.sci.osaka-u.ac.jp},
Isaac Shlosman$^{3,1}$\thanks{E-mail: shlosman@pa.uky.edu}
\\
$^{1}$ Theoretical Astrophysics, Department of Earth \& Space Science, Osaka University,
     1-1 Machikaneyama, Toyonaka, Osaka 560-0043, Japan\\
$^{2}$ Department of Physics \& Astronomy, University of Nevada Las Vegas, 4505 S. Maryland Pkwy, Las Vegas,
     NV 89154-4002, USA\\
$^{3}$ Department of Physics \& Astronomy, University of Kentucky, Lexington, KY 40506-0055, USA\\
}

\begin{document}

\date{Accepted ?; Received ??; in original form ???}


\maketitle

\begin{abstract}
We provide detailed comparison between the AMR code Enzo-2.4 and
the SPH/$N$-body code GADGET-3 in the context of isolated or cosmological
direct baryonic collapse within dark matter (DM) halos to form supermassive
black holes. Gas flow is examined by following evolution of basic parameters
of accretion flows.  Both codes show an overall agreement in the general
features of the collapse, however, many subtle differences exist.
For isolated models, the codes increase their spatial and mass resolutions
at different pace, which leads to substantially earlier collapse in SPH
than in AMR cases due to higher gravitational resolution in GADGET-3.
In cosmological runs, the AMR develops a slightly higher baryonic
resolution than SPH during halo growth via cold accretion permeated by  mergers. 
Still, both codes agree in the buildup of DM and baryonic
structures. However, with the onset of collapse, this difference in mass
and spatial resolution is amplified, so evolution of SPH models begins to
lag behind. Such a delay can have effect on formation/destruction rate of
H$_2$ due to UV background, and on basic properties of host halos.
Finally, isolated non-cosmological models in spinning halos, with spin
parameter $\lambda\sim 0.01 - 0.07$, show delayed collapse for greater
$\lambda$, but pace of this increase is faster for AMR. Within our
simulation setup, GADGET-3 requires significantly larger computational
resources than Enzo-2.4 during collapse, and needs similar resources,
during the pre-collapse, cosmological structure formation phase.
Yet it benefits from substantially higher gravitational force and
 hydrodynamic resolutions, except at the end of collapse.
\end{abstract}


\section{Introduction}
\label{sec:intro}

The origin of supermassive black holes (SMBHs) in the galactic centers
remains an unresolved problem in astrophysics. Unless the SMBHs are
primordial, a number of alternative scenarios exists. First, they could
grow from relics of the Population\,III stars \citep[e.g.,][]{Madau.Rees:01,Abel.etal:02,Bromm.Larson:04,Volonteri.Rees:05}
--- the largely pre-galactic objects whose masses appear to be comparable
or slightly higher than normal OB stars
\citep[e.g.,][see also \citealt{Bromm:13} for review]{Turk.etal:09,Hosokawa.etal:11,Wise.etal:12,
Hirano.etal:14,Fraser.etal:15}.  Second, they could form from the runaway
collapse of compact stellar clusters, subject to general relativistic effects
\citep[e.g.,][]{Zel'dovich.Podurets:65,Shapiro.Teukolsky:85}, or stellar/gasdynamics evolution
of stellar clusters \citep[e.g.,][]{Begelman.Rees:78}.
Most appealing, however, is the third alternative which involves a direct baryonic collapse to form
massive black hole seeds of $\sim 10^4-10^7\,M_\odot$, \citep[e.g.,][]{Haehnelt.Rees:93,Loeb.etal:94,
Bromm.Loeb:03,Koushiappas.etal:04,Begelman.etal:06,Volonteri.Rees:06,Begelman.etal:08,
Begelman.Shlosman:09,Milosavljevic.etal:09,Mayer.etal:10,Schleicher.etal:10,Johnson.etal:11,
Choi.etal:13,Choi.etal:15,Latif.etal:13,Prieto.etal:13,Shlosman.etal:16} which experience
a substantial growth subsequently. Depending on the
chemical composition of the collapsing gas, this can take place when suitable dark matter
(DM) halos appear in the universe. For a primordial composition, hydrogen can be
either atomic, with a cooling floor of just below $10^4$\,K, or molecular, which
is able to cool down by additional $\sim 2$ orders of magnitude. In the former case,
for the gas to collapse, the virial temperature of DM halos must exceed the cooling floor
of hydrogen, with the corresponding mass of $\sim 10^8\,M_\odot$.
In the latter case, the critical halo masses will be smaller.

Direct collapse scenario involves gravitational collapse from typical
spatial scales of $\sim 1$\,kpc, down to $\sim 1-100$ Schwarzschild radii of the seed SMBH
--- an enormous dynamic range.  Solution of the full problem requires application of the coupled
radiative hydrodynamics of the collapsing matter, MHD,
and other ingredients, and cannot be accomplished analytically. Here we limit ourselves to the
hydrodynamical and radiation part of the problem, and, in addition, assume that the collapsing gas
is optically-thin to its own produced radiation. The heavily computational approach
employed by various groups to address this and related issues involves different numerical algorithms.
These include the adaptive mesh refinement (AMR) and Smooth Particle Hydrodynamics (SPH) codes,
which are semi-Eulerian and Lagrangian, respectively.
In this work, we compare the efficiency and reliability of two representative codes,
Enzo-2.4 AMR code \citep{Bryan.Norman:97,Norman.Bryan:99,Bryan.etal:14},
and modified version of GADGET-3 SPH/$N$-body code \citep{Springel:05}.

Previous comparison between the grid and Lagrangian codes have both achieved results within $\sim 20\%$,
in the cosmological context \citep[e.g.,][]{Oshea.etal:05}, and also found significant differences.
One of the primary results that they found was that the AMR requires significantly greater
computational resources in order to achieve the same DM halo mass function as in the SPH code.
This was due to the lower force resolution in the AMR code at early times than in SPH code,
which resulted in the lack of early growth of density perturbations at high redshifts.

In the Santa Barbara Cluster Comparison project \citep{Frenk.etal:99},
different groups have simulated formation of one galaxy cluster from identical initial conditions.
In this case, for example, the central entropy profile (within the virial radius) presents a floor
in grid codes, whereas such feature is absent in the SPH codes.
These discrepancies remain largely unexplained although more recent works have improved significantly
the agreement by using the entropy-conserving version of SPH codes \citep[e.g.,][]{Springel.Hernquist:03,
Ascasibar.etal:03, Vogelsberger.etal:12}.
The ongoing AGORA project of a broad comparison between a number of grid and
Lagrangian codes of well-resolved galaxies focuses on convergence study
of various feedback processes \citep{Kim.etal:14}. However this comparison project is in its early stage.

Statistics of driven, supersonic turbulence at a high Mach number have been performed by
\citet{Price.Federrath:10} using FLASH, a widely used Eulerian grid-based AMR code, and
the SPH code PHANTOM.  Excellent agreement between these codes has been found
at similar number of particles and grid cells for the
basic statistical properties of the turbulence, such as the slope in the velocity power spectrum,
lognormal probability distribution function (PDF), the width of the PDF, etc. At the same time,
SPH code has shown a better spatial resolution. Dense structures have been resolved already
with $128^3$ SPH particles, compared to $512^3$ cells required for the same purpose in AMR.
Unfortunately, this comparison differs fundamentally from the one performed in the present work:
it is entirely non-gravitational.  As a result, the density stratification was much weaker than
one observes during gravitational collapse, and hence the evolution in their simulation did not go
as far as in our current work.

It remains unclear, whether SPH or AMR codes are more accurate in general simulation regimes.
Both methods have their shortcomings.
For example, in AMR, if the fluid moves rapidly across the mesh, the truncation
errors lead to significant departure from the same simulation if there is no bulk velocity
\citep{Tasker.etal:08}.
\citet{Hopkins:13} have studied a number of SPH modifications, in particular alternative SPH
equations of motion that guarantee conservation and improved treatment of fluid contact
discontinuities (e.g., Kelvin-Helmholz instability).
Good agreement has been reported for supersonic flows and associated strong shocks,
but the SPH suppresses mixing in subsonic, thermal pressure-dominated regimes.
\citet{Saitoh:2013aa} have also provided a new density-independent formulation of SPH
which removes the artificial surface tension.  Their method was more formally
treated in the Lagrangian formalism by \citet{Hopkins:13}.

Another important issue for classical AMR codes is the angular momentum conservation.
Unless we can anticipate the geometry of the flow and adjust accordingly the mesh so that the
fluid velocity is perpendicular to the boundary, the accuracy is seriously compromised and
subject to artificially enhanced diffusion. Finally, the refinement criteria is somewhat
arbitrary and can produce various artefacts. At the refinement boundaries, there is a
significant loss of accuracy as the refinement is necessarily discontinuous.

On the other hand, the SPH in 3-D
suffers from a poor shock resolution, noise on the scale of the smoothing kernel, and low-order accuracy
for the treatment of contact discontinuities. Hydrodynamic instabilities like the Kelvin-Helmholtz can
be suppressed \citep{Agertz.etal:07}, although recent modifications put forth by \citet{Hopkins:13}
has been claimed to correct these problems.
In real fluids, the entropy is raised in shocks because particle collisions randomize their velocities
which increase the heat and the entropy. The SPH does not capture shocks properly because entropy must
be generated locally to dissipate the small scale velocities. In order to mimic this process, and to
prevent interpenetration of the SPH particles at the shock locations, artificial viscosity is required.
The introduction of the latter usually results in an overly viscous fluid away from shocked regions.
To overcome this, the \citet{Balsara:95} switch or \citet{Cullen.Dehnen:10} formulation
have been developed.

The aim of this paper is to compare the ability of AMR and SPH codes in following the direct
collapse of gas with a primordial composition, i.e., involving identical atomic hydrogen cooling.
We compute and analyze models of gas collapse inside DM halos in isolated and cosmological settings.
Our models span more than 7 orders of magnitude in radius, from 1\,kpc down to $10^{-4}$\,pc.
Of course, the main question is whether both codes can describe the evolution that leads to the
same end product. However, specific details are important as well, namely, are
various physical quantities (e.g., gas density, temperature, tangential and radial velocity, and
angular momentum distributions) similar at various times of the collapse?
Does it take the same time to collapse,
and how do accretion rates and their temporal and radial distributions compare?

This paper is structured as follows.
In Section\,\ref{sec:method}, we briefly describe the AMR and SPH codes used.
Section\,\ref{sec:results} provides the results of our comparison runs,
and in the last section, we discuss our results and summarize them.

\section{Numerical techniques}
\label{sec:method}

\subsection{Enzo AMR and GADGET-3 SPH}
\label{sec:enzo_gadget}

The AMR code Enzo-2.4 has been tested extensively and is publicly available.
It uses a multi-grid particle-mesh $N$-body method to calculate the gravitational dynamics,
including collisionless DM particles, and a
second-order piecewise parabolic method \citep[PPM,][]{Colella.Woodward:84,Bryan.etal:95} to solve hydrodynamics.
ZEUS hydro is also available, but we use PPM here.

The structured AMR used in Enzo places no fundamental restrictions on the number of rectangular
grids used to cover some region of space at a given level of refinement,
or on a  number of refinement levels \citep{Berger.Colella:89}.
A grid is refined by a factor of 2 in lengthscale, if either the gas or
DM density become greater than $\rho_{\rm 0} N^\ell$, where $\rho_{\rm 0}$ is the
cosmic mean density for the gas or DM, respectively. The refinement factor is $N = 2$,
and $\ell$ refers to the AMR refinement level.

The Jeans length has been resolved by at least 4 cells in these simulations to satisfy the
\citet{Truelove.etal:97} requirement for resolution. For the SPH code, an equivalent criterion
is to resolve the local Jeans mass by having at least twice the number of particles in an SPH
kernel \citep{Bate.Burkert:97}.

In GADGET-3, the gas and the DM are represented by particles.
The $N$-body solver is essentially the same as in GADGET-2 \citep{Springel:05},
with some improvements for optimization purposes, such as the domain decomposition.
To achieve the spatial adaptivity at a moderate computational cost, it uses a hierarchical
multipole expansion, i.e., the tree algorithm \citep{Barnes.Hut:86}.
The gravitational potential is softened below a spatial scale specified by the gravitational
softening length.
In principle, this kernel (and the associated softening length, $\epsilon$, which represents
the force resolution) can differ from the
smoothing length used in the hydrodynamical method, as we describe below.
But mostly the SPH uses the same cubic-spline kernel, with a fixed $\epsilon$ evaluated
as the mean initial interparticle distance, divided by a fudge factor of $\sim 20-30$.

We use GADGET-3 with the density-independent SPH version with a quintic kernel and time-independent
viscosity \citep{Hopkins:13}. Comparison of various viscosity algorithms has been performed by
\citet{Cullen.Dehnen:10}, who proposed a new artificial viscosity prescription.
This viscosity grows rapidly in strong shocks, and decays rapidly away from the shocks to a
minimum $\alpha = 0.2$, compared to usual constant artificial viscosity.
With this method, the gas becomes virtually inviscid away from the shocks,
while maintaining particle order.
The viscosity decay length is taken as 3.73, and the source scaling equal to 2.77.

For the cooling package, we use Grackle version 2.0 for GADGET-3,
and Grackle version 1.1 for Enzo. These versions for the SPH and AMR codes have identical
chemistry and cooling \citep[][https://grackle.readthedocs.org/]{EnzoCollab.etal:14,Kim.etal:14}.
This library was born out of the chemistry and cooling routines
of the Enzo simulation code. The Grackle solves for
radiative cooling and internal energy, calculates cooling time, temperature, pressure,
and ratio of specific heats. It uses a non-equilibrium primordial chemistry network for atomic H and
He \citep{Abel.etal:97,Anninos.etal:97}, H$_2$ and HD, Compton cooling off the cosmic microwave
background, tabulated metal cooling and photo-heating rates calculated with Cloudy
\citep{Ferland.etal:13}. In addition, Grackle provides a look-up table for equilibrium cooling.
The gas is assumed to be dust-free and optically-thin and the metals are assumed to be in ionization
equilibrium. The cooling rate for a parcel of gas with a given density, temperature, and
metallicity, that is photoionized by incident radiation of known spectral shape and intensity can
be pre-calculated.  As we focus on gravitational collapse models at $z > 10$, and
limit the runs to the optically-thin regime, the UV background is not included, and we neglect
the radiative transfer of ionizing photons in the present work.

\subsection{Setup in Isolated Models}
\label{sec:initial-isolated}

For isolated models, we adopt the WMAP5 cosmological parameters \citep{komatsu.etal:09}, namely,
$\Omega_{\rm m} = 0.279, \Omega_{\rm b}=0.0445, h=0.701$, where $h$ is the Hubble constant
in units of $100\,{\rm km\,s^{-1}\,Mpc^{-1}}$.
We set up the details of an isolated DM halo that is consistent with the cosmological context that
we work with.  Therefore, some of the halo parameters are specified with units that include the
Hubble parameter, although we use physical quantities (not comoving) in all isolated case calculations.
In both codes, we define a DM halo as having the density equal to the critical density
$\Delta_{\rm c}$ times the mean density of the universe, $\rho_{\rm b}$, which
depends on the redshift $z$ and the cosmological model.
The top-hat model is used to calculate  $\Delta_{\rm c}(z)$, and the density
is calculated within a virial radius, $R_{\rm vir}$.
The halo virial mass is $M_{\rm vir}(z)=(4\pi/3)\Delta_{\rm c}(z)\rho_{\rm b}R_{\rm vir}$.

For isolated models we work in physical coordinates. We assume that the gas fraction
in the model is equal to the universal ratio, and simulate the gas evolution
within DM halos of a virial mass of $M_{\rm vir} = 2\times 10^8 h^{-1}\,M_\odot$ and a
virial radius
$R_{\rm vir} = 945 h^{-1}\,{\rm pc}$. The initial temperature of the gas is taken to be
$T = 3.2\times 10^4$\,K.  The simulation domain is a box with
a size $L_{\rm box}=6$\,kpc centered on the halo. In the following, we abbreviate the
spherical radii with $R$ and cylindrical ones with $r$.

\begin{figure}
  \centering
  \includegraphics[width=0.5\textwidth]{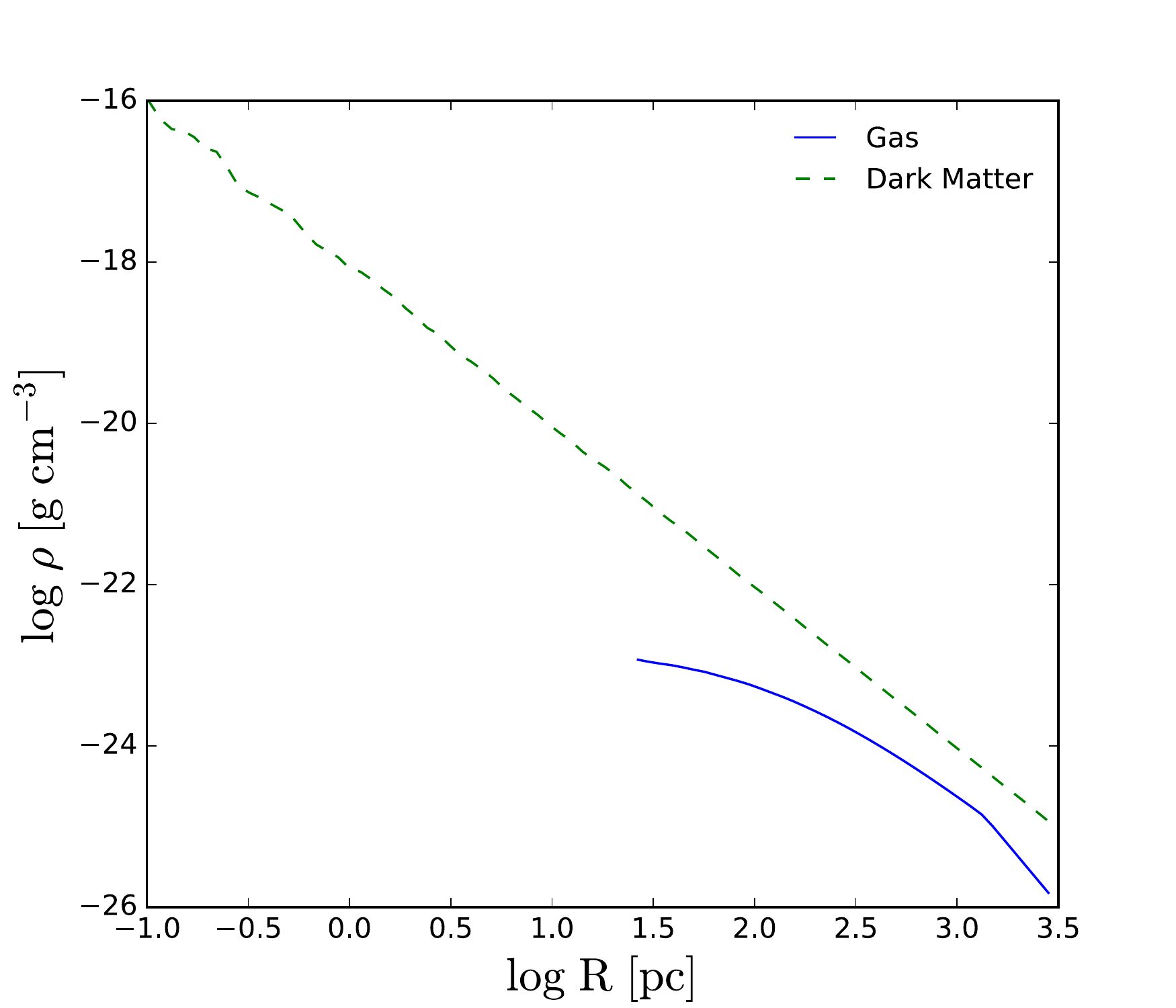}
  \caption{Initial density profiles in the isolated halo model for the gas (solid line) and DM
  (dashed line) as a function of radius.}
  \label{fig:initial_density_profile}
\end{figure}

The initial DM and gas density profiles $\rho_{\rm DM}(R)$ and $\rho_{\rm g}(R)$ are
those of an nonsingular isothermal spheres with a flat density core
of $R_{\rm c,DM} = 0.34$\,pc for DM and $R_{\rm c,g} = 142.65$\,pc for the gas,
respectively (Fig.\,\ref{fig:initial_density_profile}):
\begin{equation}
  \label{eq:1}
  \rho_g(R)=\rho_{\rm g,0} \left\{
      \begin{array}{lcl}
	\displaystyle\frac{1}{(1+\frac{R}{R_{\rm c,g}})^2}  & (R\leq R_{\rm vir}) \\
	\displaystyle\frac{1}{(1+\frac{R_{\rm vir}}{R_{\rm c,g}})^2}
                   \frac{R_{\rm vir}^3}{R^3} & (R > R_{\rm vir}),
      \end{array}
\right.
\end{equation}

\begin{equation}
  \label{eq:2}
   \rho_{\rm DM}(R)=\rho_{\rm DM,0} \displaystyle \frac{R_{\rm c,DM}^2}{R^2} \quad
       (R\leq R_{\rm vir}),
\end{equation}
where $\rho_{\rm DM,0}=7.38 \times 10^{-18}$\,g\,cm$^{-3}$,
and $\rho_{\rm g,0} =1.66 \times 10^{-23}$\,g\,cm$^{-3}$.
Such DM halos are similar to the NFW halos \citep{Navarro.etal:97}, and are simple to construct.
The DM halo core size is actually set by the gravitational softening length for GADGET-3 SPH.

The DM halo rotation is defined in terms of the cosmological spin parameter $\lambda$,

\begin{equation}
  \label{eq:3}
    \lambda=\frac{J}{\sqrt{2} M_{\rm vir} R_{\rm vir} v_{\rm c}}
\end{equation}
where $J$ is the total angular momentum of the DM halo, $M_{\rm vir}$ the halo virial mass, and
$v_{\rm c}$ the circular velocity at $R_{\rm vir}$, which is constant with radius for an
isothermal sphere mass distribution. The mean
DM spin is $\lambda\sim 0.035\pm 0.005$ \citep[e.g.,][]{Bullock.etal:01}.
We explore the range of $\lambda\simeq 0 - 0.07$ in our isolated models.
For identical $\lambda$, we sample the same initial density distribution using four different
random number sequences. The properties of the collapsing baryons have been averaged over these
sequences. Models having identical mass distribution and differing only in the initial random
sequence, are abbreviated similarly.

\begin{table}
  \centering
  \caption{Simulation parameters for the isolated halo models with Enzo AMR and GADGET-3 SPH.
  Name of the run and the values of spin parameter $\lambda$ is listed.
  ``I" in the model abbreviation stands for isolated runs, and the number $\lambda \times 100$ for DM
  halos. Cosmological models (not shown in the Table), are abbreviated by ``C'' and by $\lambda\times
  100$.}
  \begin{tabular}[!h]{ccc}
    \toprule
    Name: isolated & $\lambda$ \\
    \midrule
    AMR-I1 & 0.01  &          \\
    AMR-I3 & 0.03  &          \\
    AMR-I5 & 0.05  &          \\
    AMR-I7 & 0.07  &          \\
    \midrule
    SPH-I1 & 0.01  &          \\
    SPH-I3 & 0.03  &          \\
    SPH-I5 & 0.05  &          \\
    SPH-I7 & 0.07  &          \\
    \bottomrule
  \end{tabular}
  \label{tab:simulation_parameters}
\end{table}

To produce the DM halos with a pre-specified $\lambda$ for isolated halo models,
we follow the the prescription of \citet{Long.etal:14}.
In short, we set the isotropic distribution of velocity dispersion in DM.
Then reverse tangential velocities for a fraction of DM particles to reproduce the specific angular
momentum described above, with $\lambda$ equal to the required value.
For both SPH particles and the gas in AMR grid cells, we calculate the
average tangential velocities of the background DM in cylindrical shells, accounting for the
dependence along the (rotation) $z$-axis.
The rotational velocities for the gas in the equatorial plane of the DM halo are given by

\begin{equation}
  \label{eq:4}
  v_{\rm t}(r)=v_0 \times \left\{
      \begin{array}{lcl}
        \displaystyle r/R_{\rm c,g} & \mbox{if} & r \leq R_{\rm c,g}, \\
        1 &  \mbox{if} & R_{\rm c,g} \leq\, r\, \leq R_{\rm vir}, \\
      \end{array}
\right.
\end{equation}
where $v_0$ is tangential velocity defined by $\lambda$. The radial velocity dispersion in the
center is given by $\sigma = \sqrt{GM_{\rm vir}/2R_{\rm vir}} = 21.3\,{\rm km\,s^{-1}}$, where $G$
is the gravitational constant.
The isolated halo models have been listed in Table\,\ref{tab:simulation_parameters} and are
identified with ``I" and the value of the spin parameter multiplied by 100.
In cosmological models, we use the same notation, but replace ``I" with ``C".

In the isolated models of GADGET-3, the DM resolution is set by the fixed gravitational softening
length, $\epsilon_{\rm DM} = 0.37$\,pc. This is done in order to match the condition in Enzo runs.
For Enzo, it corresponds to an initial root grid of $64^3$ in a 6\,kpc region
with a maximal refinement level of 8 allowed for gravity, $\epsilon_{\rm DM,min} = 6000/64/2^8 =
0.37$\,pc.

For the gas, the gravitational softening is $\epsilon_{\rm g}=10^{-4}$\,pc, which is a fixed number for
GADGET-3 and serves as a minimal value for Enzo. In other words, we use the initial resolution of
$512^3$ SPH particles or cells for baryons, and $100^3$ particles or particles-in-mesh for the DM.
All other parameters in Enzo and GADGET-3 runs are similar.
The force resolution in adaptive PM codes is twice the minimal cell size
\citep[e.g.,][]{Kravtsov.etal:97}. Isolated and cosmological runs and their parameters are summarized in
Tables\,\ref{tab:simulation_parameters} and \ref{tab:setup}. We use the models AMR-I5 and SPH-I5 as
representative isolated models.

\begin{table*}
 \centering
  \caption{Summary of simulation setup for Enzo-2.4 (AMR) and GADGET-3 (SPH) simulations. }
  \begin{tabular}[!h]{lcc}
    \toprule
    Parameters & Isolated model & Cosmological model\\
    \midrule \midrule
	Simulation Volume & 6 kpc box & $3h^{-1}$Mpc with a zoom-in region of $0.4h^{-1}$\,Mpc \\
	\midrule
    Initial baryonic resolution & $64^3$ root-grid with 3 levels of refinement; & $256^3$ root-grid and SPH particles;  \\
	& Effective $512^3$ zoom grid and SPH particles & Effective $2048^3$ zoom grid and SPH particles\\
    \midrule
     Number of DM particles  & $100^3$ & $256^3$ for outer region;  \\
        &     &   Effective $2048^3$ for zoom region$^{\dagger}$\\
	\midrule
	Mass resolution        & $532.5\,M_{\odot}$ for DM particles & $245.7\,h^{-1}\,M_{\odot}$ for DM particles \\
	                       & $0.7\,M_{\odot}$ for gas particles & $44.7\,h^{-1}\,M_{\odot}$ for gas particles \\
	\midrule
	Gravitational softening & $\epsilon_{\rm DM,min}=0.37$\,pc for DM in Enzo$^{\#}$ & $\epsilon_{\rm
DM,min}=0.36h^{-1}$\,pc for DM in Enzo$^{\#\#}$  \\
                                & $\epsilon_{\rm DM}=0.37$\,pc for DM in GADGET & $\epsilon_{\rm DM}=0.36h^{-1}$\,pc for DM in GADGET\\
                                & $\epsilon_{\rm g,min}=$ smallest cell size for gas in Enzo  &
$\epsilon_{\rm g,min}=$ smallest cell size for gas in Enzo \\
		 & $\epsilon_{\rm g}=10^{-4}$\,pc for gas in GADGET & $\epsilon_{\rm g}=10^{-3}h^{-1}$\,pc for gas in GADGET \\
	\midrule
    Minimum gas smoothing length $\eta_{\rm min}$ & $10^{-3}$\,pc & $10^{-2}h^{-1}$\,pc $^{\#\#\#}$ \\
    \bottomrule
  \end{tabular}
$^\dagger$ The size of the zoom-in region is about one-tenth of the outer region, therefore the actual grid/particle number in the zoom region is initially $\sim 200^3$ for both baryons and DM particles. \\
$^\#$ Maximum refinement level for DM gravity is set to 8 levels for Enzo in isolated models, i.e., $\epsilon_{\rm
DM,min}=6000/64/2^8=0.37$\,pc.  The isolated models do not have Hubble
expansion in space, so there is no factor of $h^{-1}$ and all numbers are physical.  \\
$^{\#\#}$ Maximum refinement level for DM gravity is set to 15 levels for Enzo in cosmological models, i.e., $\epsilon_{\rm
DM,min}=3\times 10^6/256/2^{15}=0.36h^{-1}$\,pc.  \\
 $^{\#\#\#}$ Maximum refinement levels for the gas in Enzo are set to 17 levels in isolated models and 20 levels in cosmological models. \\
  \label{tab:setup}
\end{table*}

\subsection{Setup in Cosmological Models}
\label{sec:initial-cosmological}

For further comparison, we use zoom-in cosmological simulations with the same
composition as in isolated models to follow up the gas evolution within DM halos.
The initial conditions (ICs) are generated using WMAP5 cosmology:
$\Omega_{\rm \Lambda} = 0.721$, $\Omega_{\rm m} = 0.279$, $\Omega_{\rm b}
= 0.0445$, $h=0.701$, $\sigma_8 = 0.807$, and $n = 0.961$. The models run
from $z=200$.
The same cooling packages have been used for cosmological models as for isolated
ones (Section\,\ref{sec:enzo_gadget}).

For the initial setup, we use the MUSIC algorithm \citep{Hahn.Abel:11}
to generate cosmological zoom-in ICs.
MUSIC uses a real-space convolution approach
in conjunction with an adaptive multi-grid Poisson solver to
generate highly accurate nested density, particle displacement,
and velocity fields suitable for multi-scale zoom-in
simulations of structure formation in the universe.

Generating a set of zoom-in ICs is a two-step process.
First, we generate $3h^{-1}$\,Mpc comoving $256^3$ DM-only ICs for the
pathfinder simulation and run it without AMR until $z = 10$.
Using the HOP group finder \citep{Eisenstein.Hut:98}, we select an appropriate DM halo,
whose mass is $\gtorder 10^8 h^{-1}\,{\rm M_\odot}$ at $z = 10$.
Second, we generate $0.4h^{-1}\,{\rm Mpc}$ ICs with $2048^3$ effective resolution
in DM and gas, embedded in the lower resolution outer region.
Since we use the same random seeds for these ICs as the first step,
the phases of both ICs are identical. The zoom-in region is centered on the selected
halo position and is set to be large enough to cover the initial positions of all
selected halo particles. We perform the zoom-in procedure for each of the
targeted halos.

In the cosmological runs, the gravitational softening is $10^{-3}$\,pc
in the comoving coordinates, in the SPH run.
We have measured the spin parameter of many halos in the range of $\lambda\sim 0.01-0.07$
in our zoom region at $z\sim 10$,
and use $\lambda\sim 0.04$ as a representative one for both AMR and SPH runs.

\section{Results}
\label{sec:results}

We aim at comparing results for direct baryonic collapse within DM halos by Enzo-2.4 AMR
and GADGET-3 SPH codes. As a first step, we compare the isolated models.
The cosmological halos have a range of spin parameter, as discussed in
\S\,\ref{sec:initial-isolated}.
To match the conditions of these halos, we generate isolated halos with the same range of
$\lambda$ (Table\,\ref{tab:simulation_parameters} and \S\,\ref{sec:res_isolate}).
The cosmological halos are examined in \S\,\ref{sec:res_cosmo}.
Lastly, we test the effect of $\lambda$ on the dynamics of the collapse in isolated and
cosmological models (\S\S\,\ref{sec:res_spin},\ref{sec:res_cosmo_spin}).

\subsection{Direct collapse in isolated models}
\label{sec:res_isolate}

\begin{figure}
  \centering
  \includegraphics[width=0.46\textwidth]{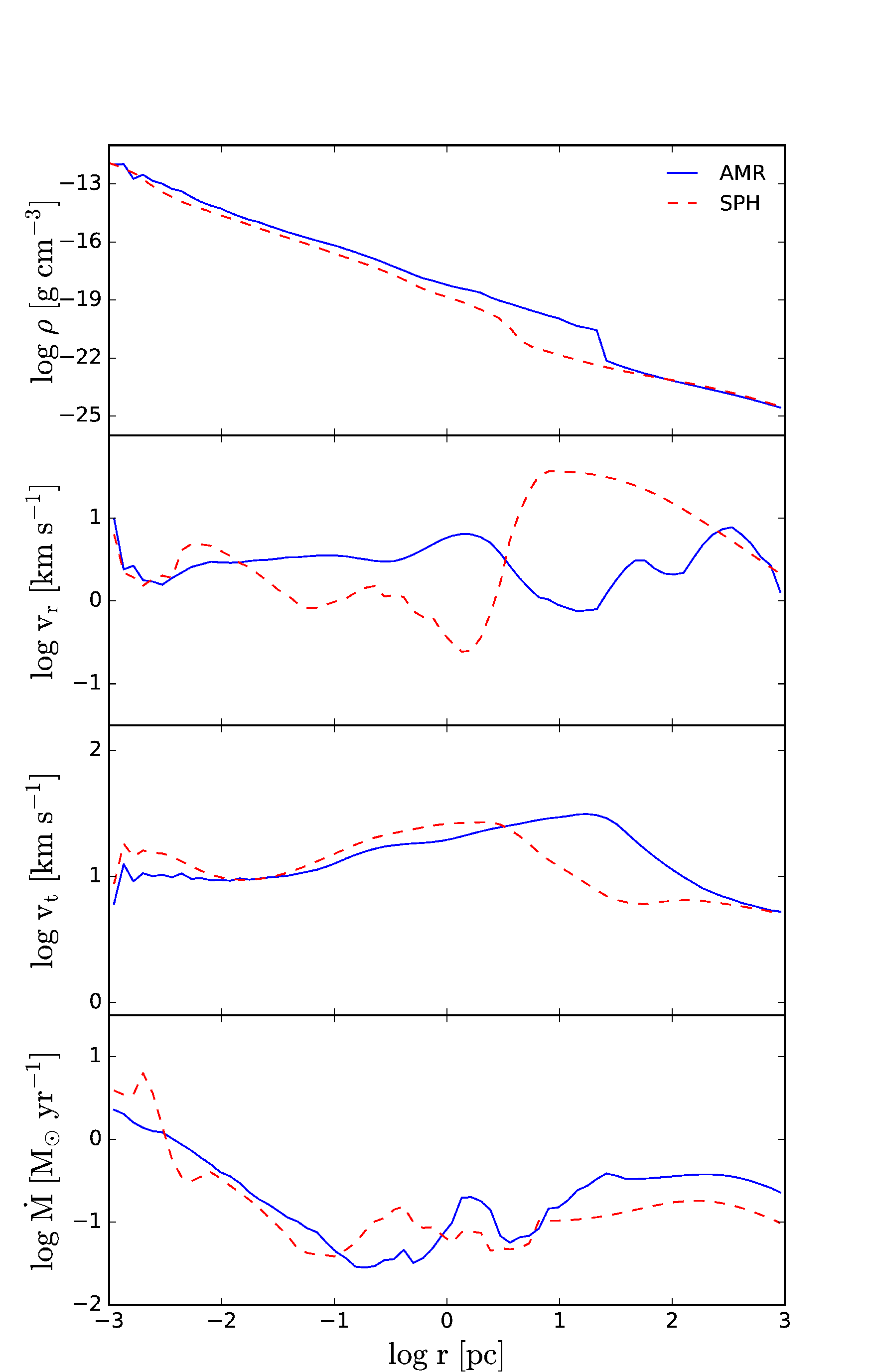}
\caption{From top to bottom, radial profiles of gas density, radial velocity, tangential velocity,
and mass accretion rate, at the end of the isolated runs for models AMR-I5 (solid lines) and
SPH-I5 (dashed lines).  All axes are logarithmic.  The density and accretion rates have been
averaged over spherical shells, and the velocities averaged over cylindrical shells.
}
  \label{fig:comparison_dynamics}
\end{figure}

\begin{figure}
  \centering
  \includegraphics[width=0.52\textwidth]{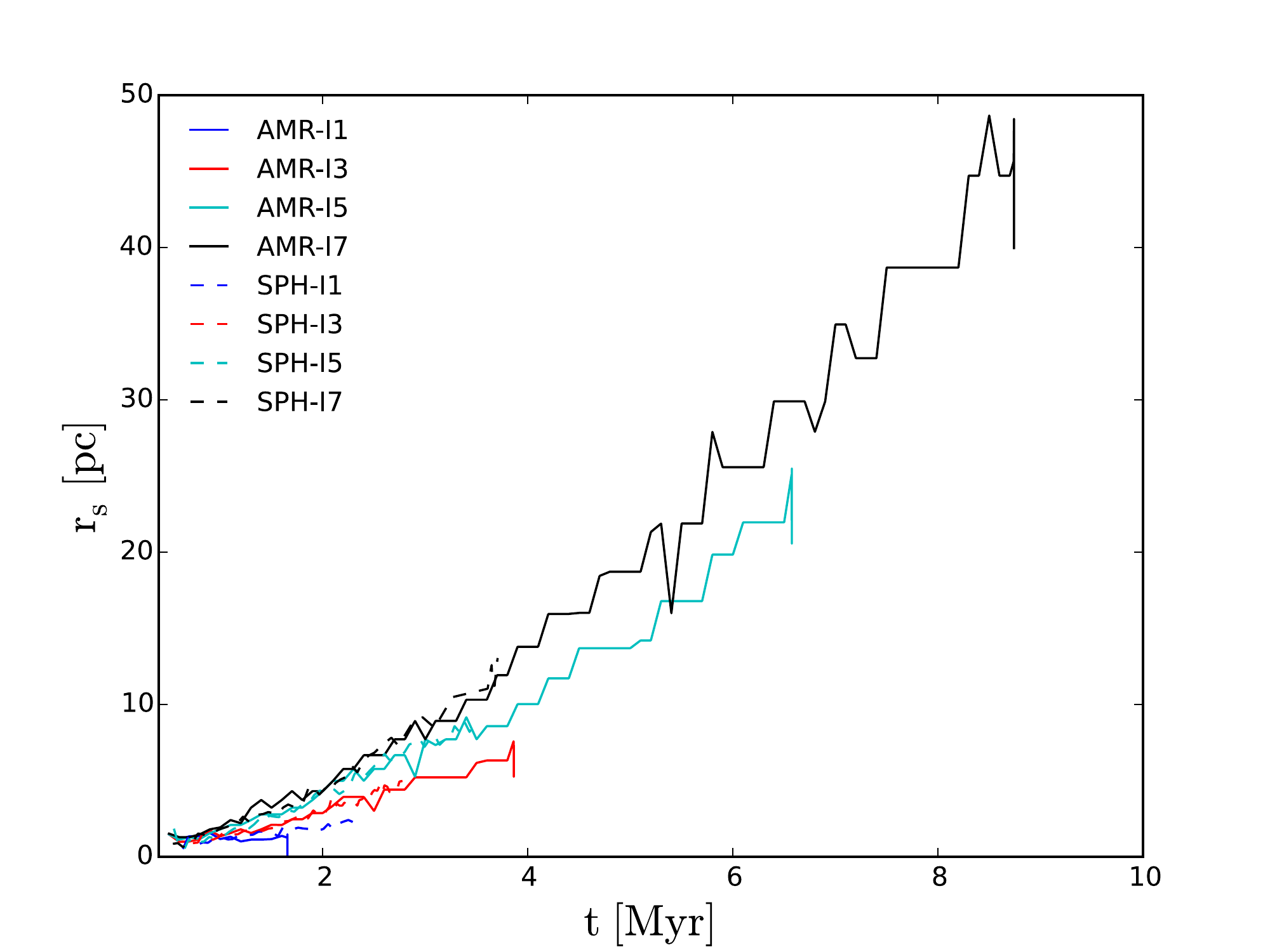}
\caption{Evolution of the radial position of shocks, $r_{\rm s}$, in isolated models in Enzo (solid
lines) and GADGET-3 (dashed lines) runs for various $\lambda$. Note that both AMR and SPH runs with
the same $\lambda$ exhibit the same $r_{\rm s}(t)$. The maximal $r_{\rm s}$ correspond to the end
of the simulation.
}
  \label{fig:shock_evolution}
\end{figure}

Direct collapse of isolated models with atomic gas has been performed and analysed
by \citet{Choi.etal:13} for $\lambda = 0.05$, using Enzo.
Many of the aspects of its evolution are generic.
We define $t=0$ at the start of the run, and stop the run when the central collapse reaches
the final resolution of
$R_{\rm fin}\sim 10^{-3}$\,pc, in both Enzo and GADGET-3, in order to allow a meaningful
comparison, because the SPH run would have difficulty reaching a higher resolution without
invoking complicated procedures, such as SPH particle splitting.

The DM in isolated models has isothermal profiles by construction, i.e., $\rho_{\rm DM}
\propto R^{-2}$. They remain nearly identical to ICs.
As the collapse proceeds and reaches the central region, the DM is dragged inward by the gas,
and is compressed adiabatically, forming new cusps with slopes $\sim -1$ to $\sim -2$.

The baryonic collapse proceeds in two stages, which are representative for all isolated models.
First, the gas cools down to the cooling floor of atomic gas, then collapses following
a self-similar solution, analogous to the Larson-Penston solution for an isothermal, self-gravitating
gas cloud, $\rho \propto R^{-2}$ \citep{Larson:69,Penston:69}. But significant differences are obvious,
as angular momentum is present and modifies the solution, and the DM dominates the gravitational
potential everywhere during the initial stage.

When collapsing gas reaches the centrifugal barrier, a shock forms, where the density jumps
and the radial velocity drops abruptly (top panels of Figure\,\ref{fig:comparison_dynamics}).
The gas accumulates behind the shock and forms a disk which grows in mass.
As the gas density in the inner disk surpasses that of the DM, the 2nd stage of the collapse ensues,
and proceeds with an increasingly short timescale.
Basically, most of the gas behind the shock collapses and reaches the prescribed resolution of
$\sim 10^{-4}$\,pc \citep[e.g.,][]{Choi.etal:13}, although here we compare the final
distributions when $R_{\rm fin}\sim 10^{-3}$\,pc has been reached.

The current AMR simulations with various $\lambda$ follow basically the same outline. The initial
positions of shocks in all runs are located  at $\sim 1$\,pc, and move out with time.
The radial positions of these shocks also increase with increasing spin parameter, $\lambda$,
as the centrifugal barrier is reached at progressively larger radii.
Figure\,\ref{fig:shock_evolution} displays the shock locations as function of time,
from their formation time to the end of the simulations.  Note, that for a
specific $\lambda$, both AMR and SPH shocks strictly follow each other. Moreover, consistently, the
AMR shocks reach larger radii, as shown in Figure\,\ref{fig:shock_evolution} for all $\lambda$.
The explanation for this difference comes from the fact that the isolated AMR models
collapse later than the isolated SPH models, as we discuss below, and shocks have more time
to advance.

As a first step, we compare the radial profiles of a gas density, radial and tangential velocities,
and accretion rates at the end of the runs, i.e., when the collapse has reached $R_{\rm fin}$
(Fig.\,\ref{fig:comparison_dynamics}).  All values are given after averaging in cylindrical or
spherical shells
and at the run end. The gas density and accretion rates have been
averaged over spherical shells, and the velocities averaged over cylindrical shells.
Shown are the representative AMR-I5 and SPH-I5 runs.
In AMR-I5 run, the gas goes through a strong accretion shock and forms a disk,
ending the first stage of the collapse. By the end of the runs,
the SPH-I5 shock is positioned at a radius about 3 times smaller than in the AMR-I5, which is
confirmed by Figure\,\ref{fig:shock_evolution}. It appears also somewhat weaker than the AMR shock.
In all other respects, the gas density profiles are very similar, albeit it takes another decade in
radius for the density in SPH-I5 to catch up with the AMR-I5 density just inside the shock.
The final density is $\rho\sim 10^{-14}\,{\rm g\,cm^{-3}}$ at $R_{\rm fin}=10^{-3}$\,pc.

\begin{figure*}
  \centering
\includegraphics[width=0.9\textwidth]{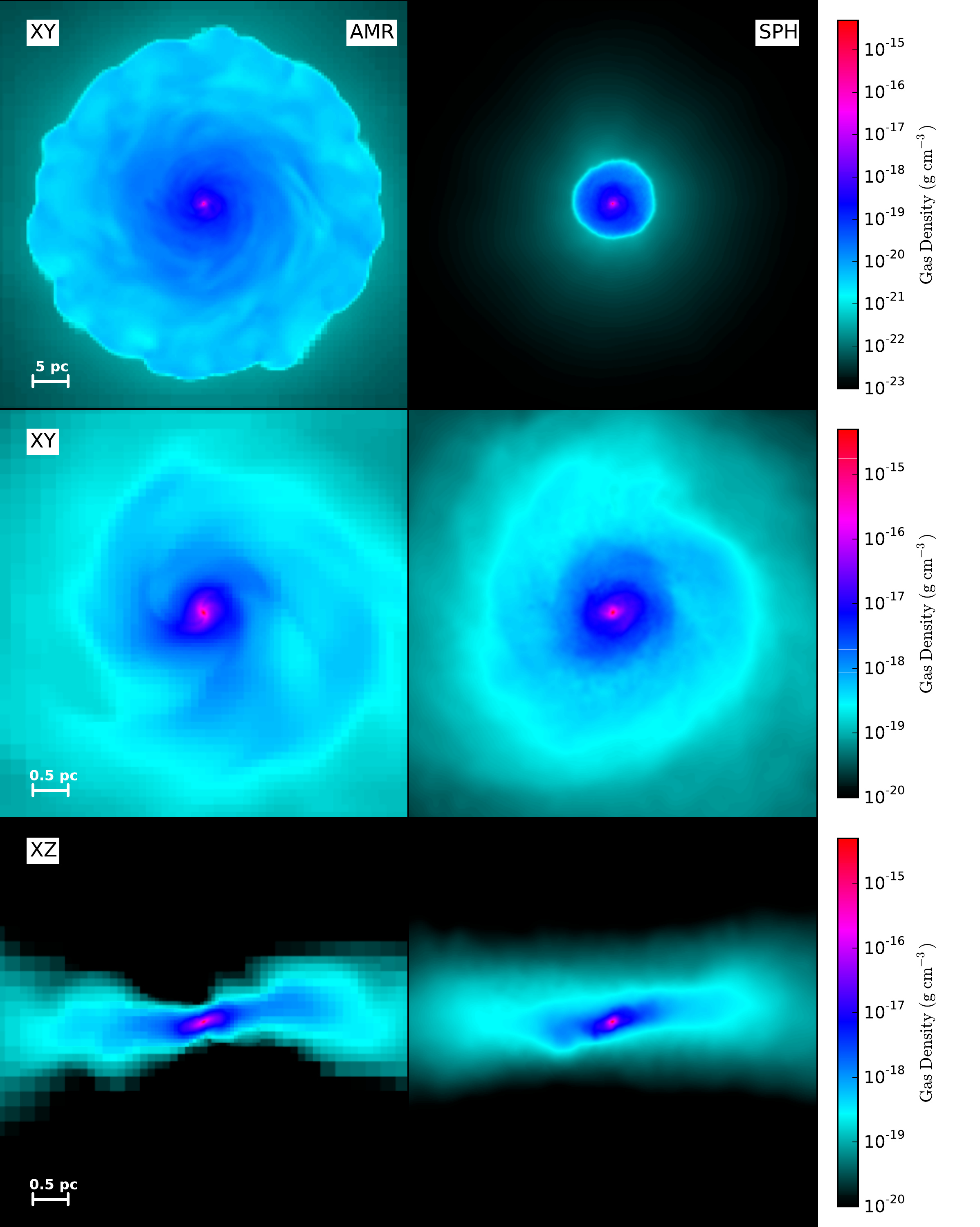}
\caption{Mass-weighted gas volume density in the $x$-$y$ equatorial plane (top and middle frames)
and the $x$-$z$ plane (bottom) of isolated models AMR-I5 (left frames) and SPH-I5 (right frames),
at the end of the runs.
Shown are the snapshots of decreasing spatial scales (physical):
50\,pc (top) and 5\,pc (middle and bottom) on each side, respectively.
The color palette reflects the range from minimum to maximum densities in each panel.
The face-on and edge-on disks can be seen clearly in both runs.
}
  \label{fig:projectiondensity}
\end{figure*}

\begin{figure*}
  \centering
 \includegraphics[width=0.85\textwidth]{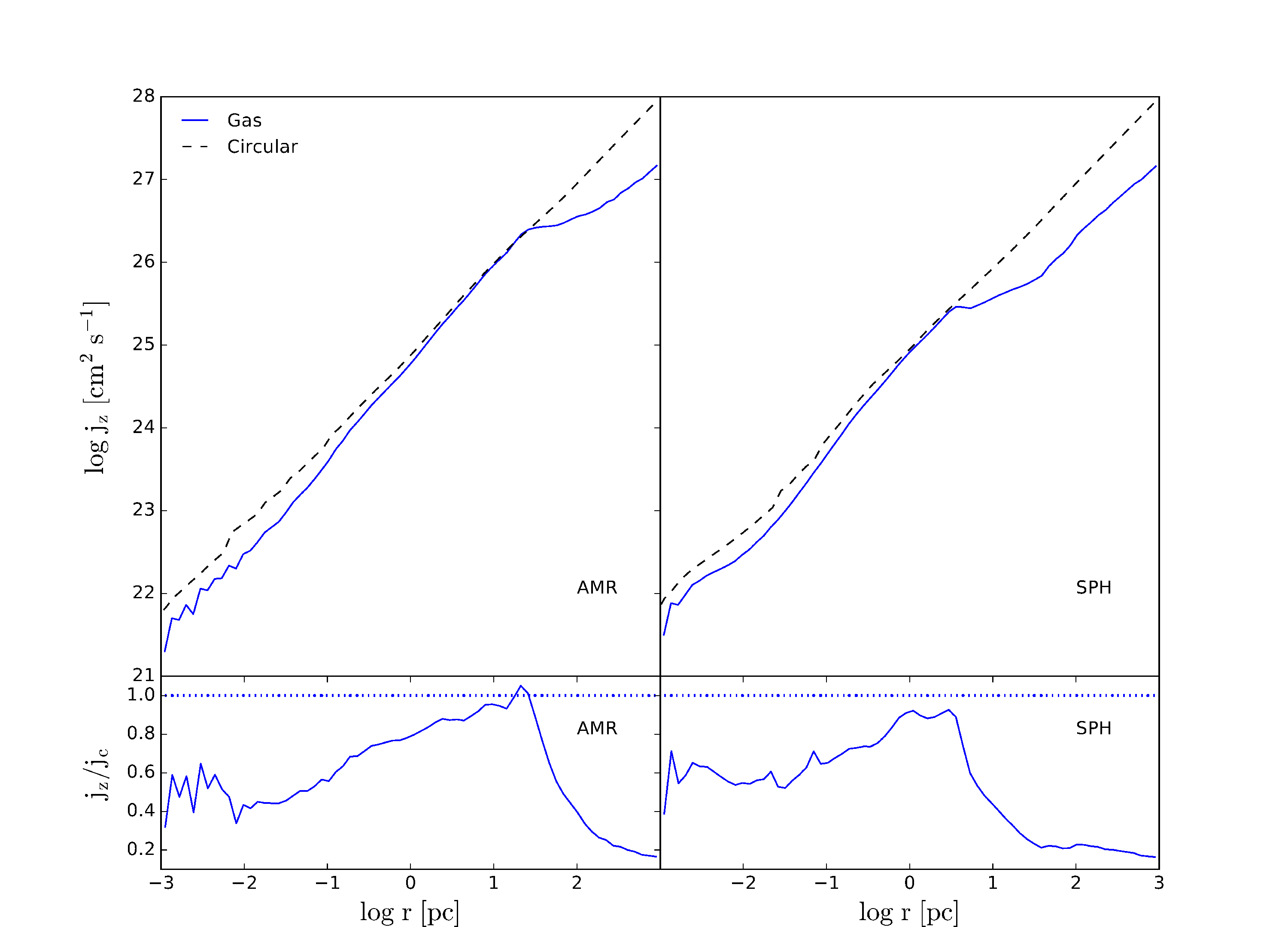}
  \caption{Specific angular momentum profile in the gas as a function of cylindrical radius
     for AMR-I5 (left panel) and SPH-I5 (right panel). Circular angular momentum profile is
shown for comparison (dashed line).}
  \label{fig:jz}
\end{figure*}

 \begin{figure*}
  \centering
  \includegraphics[width=1.0\textwidth]{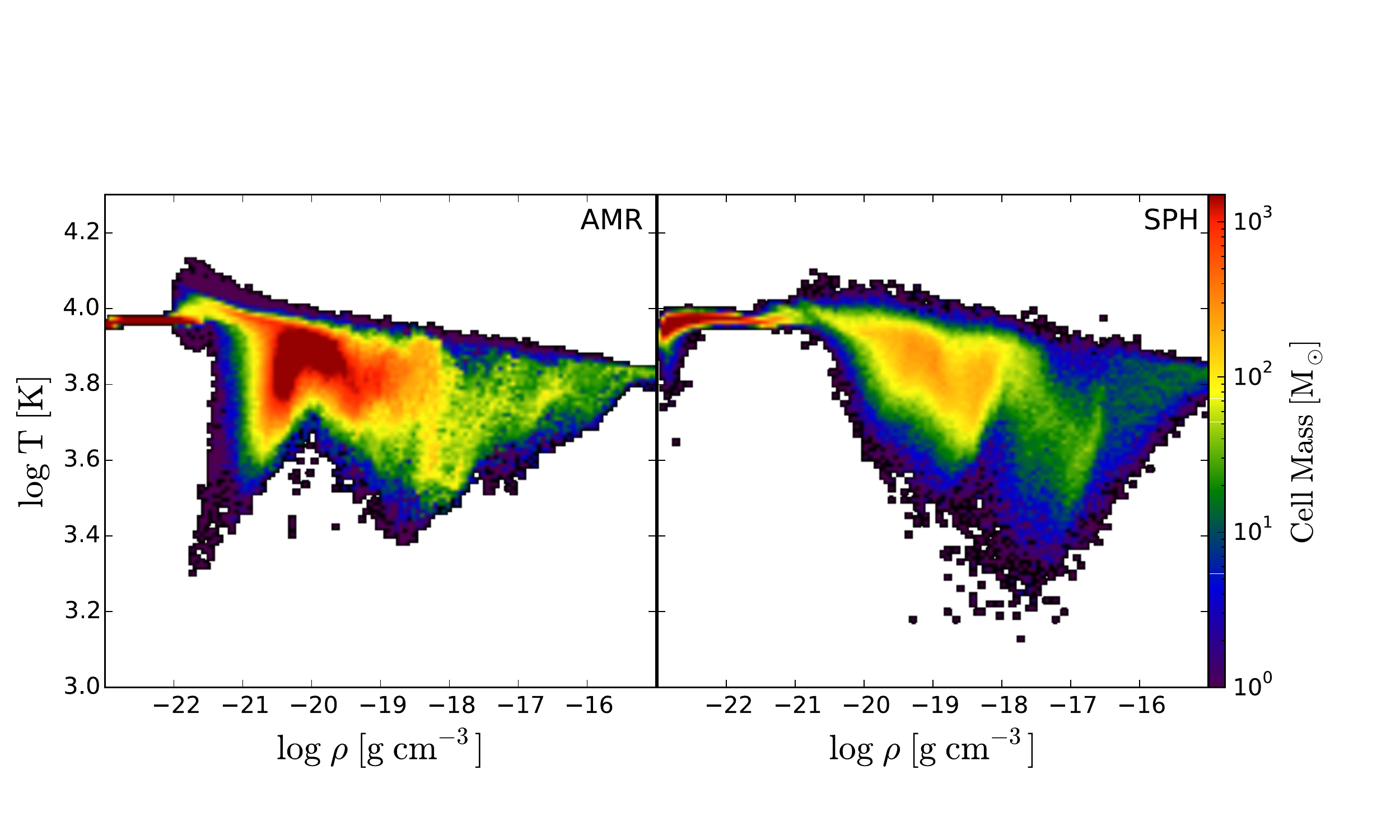}
  \caption{Phase diagram of temperature vs. density of gas in isolated runs AMR-I5 (left panel)
  and SPH-I5 (right panel) at the end of each run. The color represents the amount of mass
  in each pixel. }
  \label{fig:density_temperature}
\end{figure*}

The radial velocity profiles, $v_{\rm r}$, are very similar in the outer few hundred pc.
In the preshock region, both $v_{\rm r}$ increase toward the center,
then fall down substantially to a deep minimum in the post-shock region,
at the respective positions of the shocks in the AMR and SPH runs.
At smaller radii of $r < 10^{-2}$\,pc, the curves are very similar and increase by
an order of magnitude toward $R_{\rm fin}$.
The tangential velocities, $v_{\rm t}$, reach maximal values in the post-shock region of both runs,
as expected, then decrease toward the center.  The behavior of $v_{\rm r}$ and $v_{\rm t}$ with $r$
is associated with a variable degree of rotational support with radius, as we discuss below.

The mass accretion rate profiles show a similar behavior, but $\dot M$ differs by a factor of 3
in the preshock region, where $\dot M\sim 0.6\,M_\odot\,{\rm yr^{-1}}$ for the AMR run
and $\sim 0.2\,M_\odot\,{\rm yr^{-1}}$ for the SPH run.
This rate drops to below $\sim 0.1\,M_\odot\,{\rm yr^{-1}}$ at $R<10$\,pc for both runs.
The 2nd stage of the collapse inside $\sim 0.1$\,pc displays a nearly identical behavior
--- both curves show a linear growth with decreasing $r$,
reaching $\dot M\sim 0.5\,M_\odot\,{\rm yr^{-1}}$ at $R_{\rm fin}$. The small spike in the
SPH run at $r\sim 10^{-3}$\,pc is caused by insufficient mass resolution there.

\begin{figure*}
  \centering
\includegraphics[width=1.0\textwidth]{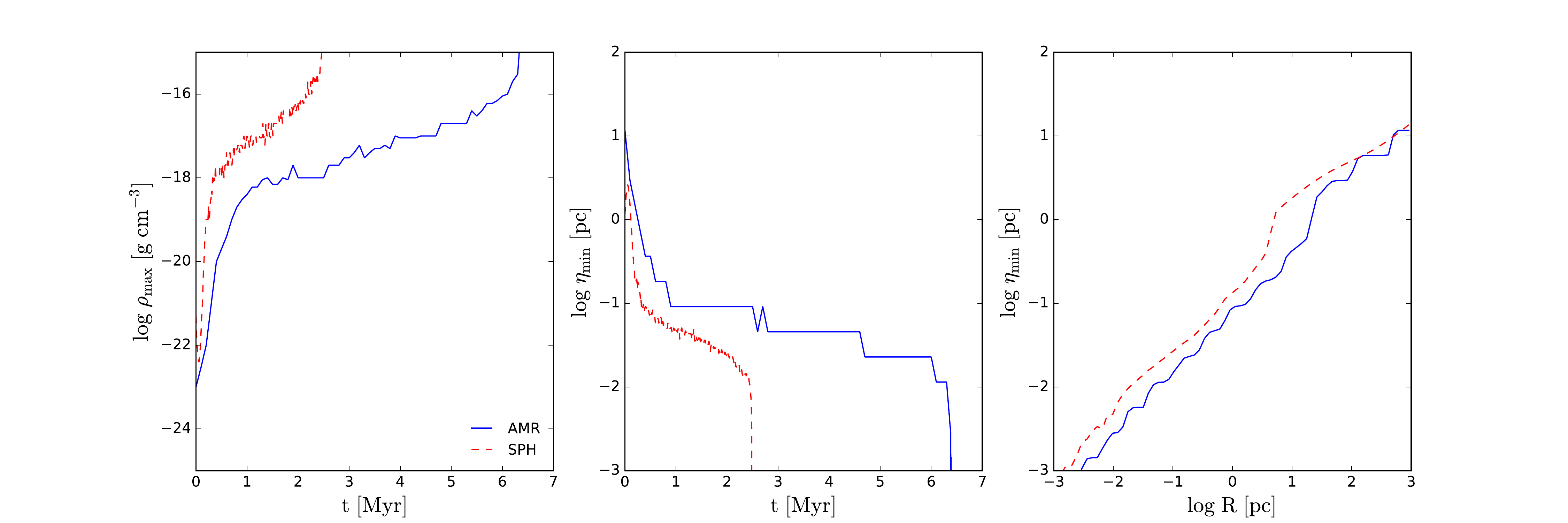}
  \caption{Comparing various resolution parameters for isolated models AMR-I5 (solid lines)
  and SPH-I5 (dashed lines).
  {\it Left panel:} Evolution of maximum gas density $\rho_{\rm max}$ as a function of time.
  {\it Middle panel:} Evolution of the minimal gas smoothing length for the SPH and minimum
  cell size for the AMR run, $\eta_{\rm min}$, as functions of time.
  {\it Right panel:} Radial profile of resolution given by the smoothing length for SPH and the
  minimum cell size at {\it each} radius for AMR, at the end of the runs. The 1st stage of the
  collapse which starts at $t=0$ is identical for both models. The 2nd stage of the collapse is
  triggered
  when the gas-to-DM density ratio in the inner few pc exceeds unity, i.e., $\sim 2.4$\,Myrs
  for GADGET and $\sim 6.3$\,Myr for Enzo.
  Note, that after the onset of the 2nd stage of the collapse, the collapse proceeds on
  a very short timescale as observed by the sharp drop in $\eta_{\rm min}$ (middle panel).
}
  \label{fig:compare_res}
\end{figure*}

The projected gas volume density shown in Fig.\,\ref{fig:projectiondensity}
emphasizes both similarities and differences, in comparison with the 1-D radial density profiles.
The most significant difference is that of the central disk size in the top frames
--- disk radius of $\sim 20$\,pc in the AMR-I5, and only $\sim 7$\,pc in the SPH-I5.
This difference clearly has its origin in the position of the radial shock in both runs,
as discussed above. So both disks appear shock-bounded.
In the top frames, we also see a pronounced spiral structure in the AMR run,
and a weaker one in the SPH run.
The same conclusion can be reached from 5\,pc scale images (middle and bottom panels),
where $m=2$ (gaseous bar) is seen within the central 0.3\,pc, and $m=3$ mode is driving a triple
spiral on scales $\gtorder 0.5$\,pc.
The SPH-I5 run displays comparable $m=2$ and 3 modes on the same scales.

Redistribution of the angular momentum is an important ingredient of gravitational
collapse. We compare specific angular momenta, $j_{\rm z}$, in the gas at the end of the
simulations, in AMR-I5 and SPH-I5 (Fig.\,\ref{fig:jz}).
Two important differences can be observed.
Positions of radial shocks (Fig.\,\ref{fig:comparison_dynamics})
corresponds to the radii where the specific angular momenta curves come close to the
circular angular momenta, $j_{\rm z}\sim j_{\rm c}$. At smaller radii, $j_{\rm z}$ moves
away from $j_{\rm c}$ gradually. Still, there is a non-negligible rotational support at
$R_{\rm fin}$, in both runs, albeit smaller for the SPH-I5 run. The bottom frames of this
Figure provide the quantitative measure of rotational support in the form of
$j_{\rm z}/j_{\rm c}$ radial profiles. The AMR exhibits a disk at larger radii than the SPH run,
because it takes more time for the AMR model to collapse, and, therefore, the accretion shock
has more time to propagate outwards. In the 2nd stage of the collapse, i.e., inside
$\sim 1$\,pc, the $j_{\rm z}/j_{\rm c}$ ratio declines inwards, and showing a plateau
for $\ltorder 10^{-2}$\,pc, at about 0.5. They demonstrate that, in the isolated models,
rotational support for the collapsing gas never falls below $\sim 50\%$ in the 2nd stage,
in both runs.

The $T-\rho_{\rm g}$ relation is shown in Figure\,\ref{fig:density_temperature} at the end
of the representative AMR-I5 and SPH-I5 runs.
While the upper envelope of $T$ is similar in both models,
there appears to be much more gas at lower $T$ at low densities,
$\rho<10^{-20}$\,g\,cm$^{-3}$, in the AMR-I5 run.
Given our refinement criteria, this is an anticipated property of mesh codes which
are able to follow
the low-density gas, while SPH codes tend to resolve better
the higher density regions.
On the other hand, a larger  amount of low $T$ gas appears to be present at high densities,
$\sim 10^{-18} - 10^{-17}$\,g\,cm$^{-3}$, in the SPH-I5 run.
We note that AMR run has larger mass accumulated (at $\rho\sim 10^{-20}\,{\rm g\,cm^{-3}}$)
around the shock, due to a larger shock radius, as evident from Figure\,\ref{fig:shock_evolution}.
Since we use identical cooling packages for all runs,
this difference can arise from the ability of numerical schemes to resolve the
shocks and the cooling of a post-shock gas.
In the AMR-I5, the radial shock is strong and advances to larger radii
corresponding to lower gas densities of $\sim 10^{-21}\,{\rm g\,cm^{-3}}$.
The low $T$ gas is formed from the post-shock gas which cools down more efficiently.
On the other hand, the SPH-I5 shock is weaker and happens at $\sim 10^{-19}\,{\rm g\,cm^{-3}}$,
where particles at lower $T$ already are much more common.
Hence the shocked particles' contribution is hardly visible on the phase diagram.

Figure\,\ref{fig:compare_res} compares various parameters which affect the resolution in AMR
and SPH runs. The left panel shows evolution of maximal density in the gas, $\rho_{\rm max}$,
with time. Clearly, the collapse in the isolated model SPH-I5 proceeds much faster than in AMR-I5
already during the 1st stage. At the onset of 2nd stage, the density increases slowly, exhibiting
kind of a plateau, then accelerates, and at the end of this stage $\rho_{\rm max}$ increases rapidly.
Both runs display an identical behavior at the end.

The evolution of $\rho_{\rm max}$ allows us to quantify
the separation of the two stages of the collapse \citep[see also][]{Choi.etal:13}.
The plateau between the stages provides an independent quantitative support to the existence
of these stages.
They are also observed in the evolution of the minimum gas smoothing length for SPH
and minimum cell size for AMR, $\eta_{\rm min}$,
shown in the middle frame of Figure\,\ref{fig:compare_res} --- this frame describes the same
process of $\rho_{\rm max}$ but in terms of the gas minimal smoothing length $\eta_{\rm min}$,
which is the minimal interparticle distance
Following \citet{Choi.etal:13,Choi.etal:15} and \citet{Shlosman.etal:16}, we choose to define
the beginning of the 2nd stage when the gas-to-DM density ratio becomes larger than unity within the
central few pc.
This happens at $t\sim 2.4$\,Myr in SPH-I5 and $\sim 6.3$\,Myr in AMR-I5.
The collapsing gas reaches the final resolution of $\eta_{\rm min}\sim {\rm few}\times 10^{-3}$\,pc
at $\sim 2.4$\,Myr and $\sim 6.3$\,Myr, respectively.
At this time the collapse has reached  $r_{\rm fin}$.

The final radial profile of $\eta_{\rm min}$ is shown in the right panel of
Figure\,\ref{fig:compare_res}.
Both AMR and SPH curves for $\eta_{\rm min}$ are similar, but still the AMR one is systematically
lower by a factor of $2-3$ depending on the radius.

\subsection{Direct collapse in cosmological models}
\label{sec:res_cosmo}

For cosmological models,
we stop the runs when the AMR and SPH resolutions, measured by $\eta_{\rm min}$,
reach $R_{\rm fin}\sim 10^{-2}$\,pc. The reason for this is that high resolution of
isolated models is more difficult to reach in cosmological runs. Therefore, we reduce
the resolution demands in both AMR and SPH cosmological models by a factor of 10 compared to
the isolated ones. The time is measured from the Big Bang.

As the DM halos form and grow with time in the cosmological models, the DM density profiles
tend to the NFW shape \citep{Navarro.etal:97}. When the collapse reaches the central regions,
the baryon drag in the DM and form a cusp, similar to that in isolated models. Note,
that our isolated models follow the isothermal sphere density profiles and not the NFW ones.
We shall return to this issue in \S\,\ref{sec:discuss}.

\begin{figure}
  \centering
  \includegraphics[width=0.5\textwidth]{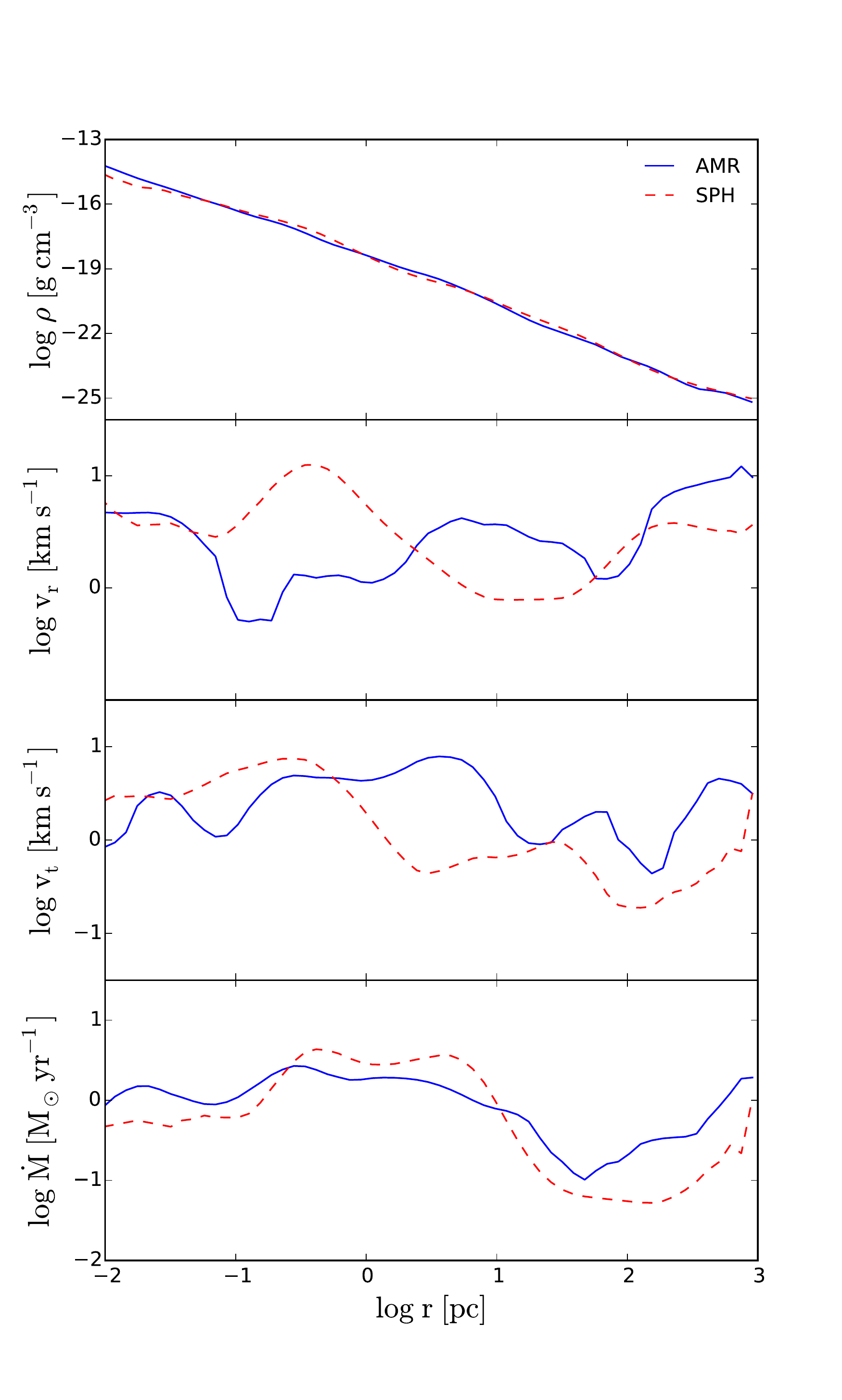}
\caption{From top to bottom, radial profiles of gas density, radial velocity, tangential
velocity, and mass accretion rate, at the end of the cosmological runs for models AMR-C4.2
(solid lines) and SPH-C4.2 (dashed lines). All axes are logarithmic. The density and
accretion rate have been averaged over
spherical shells, and velocities averaged over cylindrical shells.
}
\label{fig:cosmo_comparison_dynamics}
\end{figure}

We show the gas density, radial and tangential velocity, and mass accretion rate profiles
for a representative cosmological models AMR-C4.2 and SPH-C4.2 with $\lambda\sim 0.042$ ---
very similar to the representative isolated cases (Fig.\,\ref{fig:cosmo_comparison_dynamics}). 
We have also measured the DM halo virial masses and virial radii for Enzo and GADGET
cosmological models at the time of direct collapse, $z_{\rm coll}$
(Table\,\ref{tab:halo_at_collapse}). Overall, the AMR models collapse slightly earlier than
the SPH ones. Consequently, the DM halos appear slightly less massive and their virial radii
slightly smaller. So a pronounced trend exists between the AMR and SPH cosmological models.

\begin{table}
  \centering
  \caption{DM halo virial parameters in AMR and SPH cosmological models at the time of
  direct collapse, i.e., at $z_{\rm coll}$.
  }
  \begin{tabular}[!h]{cccc}
    \toprule
    Models   & $z_{\rm coll}$ & $M_{\rm vir}\,[h^{-1}M_\odot]$ & $R_{\rm vir}$\,[$h^{-1}$pc]  \\
    \midrule
    AMR-C1.5 & 17.2 & $2.4\times 10^7$ & 407 \\
    AMR-C2.8 & 11.9 & $4.7\times 10^7$ & 722 \\
    AMR-C4.2 & 19.3 & $2.7\times 10^7$ & 379 \\
    SPH-C1.5 & 15.5 & $4.1\times 10^7$ & 538 \\
    SPH-C2.8 & 11.0 & $5.4\times 10^7$ & 809 \\
    SPH-C4.2 & 18.6 & $3.4\times 10^7$ & 422 \\
    \bottomrule
  \end{tabular}
  \label{tab:halo_at_collapse}
\end{table}

The top panel displays the final density profiles which are remarkably similar for the two runs.
While density at $R_{\rm fin}\sim 10^{-2}$\,pc is nearly identical to that in the corresponding
isolated runs, the overall shape of $\rho(R)$ somewhat differs. In the isolated runs,
inside the radial shock, the slope of $\rho(R)$ was nearly constant, while in the cosmological runs
it exhibits the trend of becoming shallower with $R$. This change happens at $R\sim
0.5-5$\,pc, continues inward, and appears slightly more pronounced in AMR than in the SPH runs.
It appears to be the result of our use of an isothermal density profile for the DM in isolated
models, and the formation of less cuspy NFW DM profiles in the cosmological runs.
Another major difference between isolated and cosmological models is the absence
of standing radial shock, which we explain below.

The radial velocities in AMR-C4.2 and SPH-C4.2 vary within a factor of a few along $r$, but end
up similar within a factor of 2 at the center.
The tangential velocity is consistently higher in AMR at $r\gtrsim 0.2$\,pc, then reverse
the trend with the AMR velocity becomes smaller than SPH one at $r\lesssim 0.2$\,pc.

The bottom panel of Figure\,\ref{fig:cosmo_comparison_dynamics} displays the radial profile of
mass accretion rate, $\dot M$. The curves differ by a factor $\sim 2-3$ among themselves, in the
outer radii, and $\ltorder 2$ inside 50\,pc.
But they also differ profoundly from their isolated counterparts. First they exhibit a deep
minimum at $r\sim 50-100$\,pc and stay about constant, $\dot M\sim 1-3\,{\rm M_\odot\,yr^{-1}}$
inward. The reason for this behavior is increased rotational support they experience in this
region, i.e., $v_{\rm r}$ and $v_{\rm t}$ have local minima and maxima there.
On the other hand, the isolated models display growing $\dot M$ toward $r_{\rm fin}$ and reach
a maximum there which is greater than the cosmological case by a factor of 2.
The cosmological $\dot M$ is higher by a factor of a few everywhere except at the very center.
Lastly, at the center, for $R\ltorder 0.1$\,pc, the mass accretion in the SPH run falls
by almost an order of magnitude, and less so in the AMR run.

\begin{figure*}
  \centering
\includegraphics[width=0.85\textwidth]{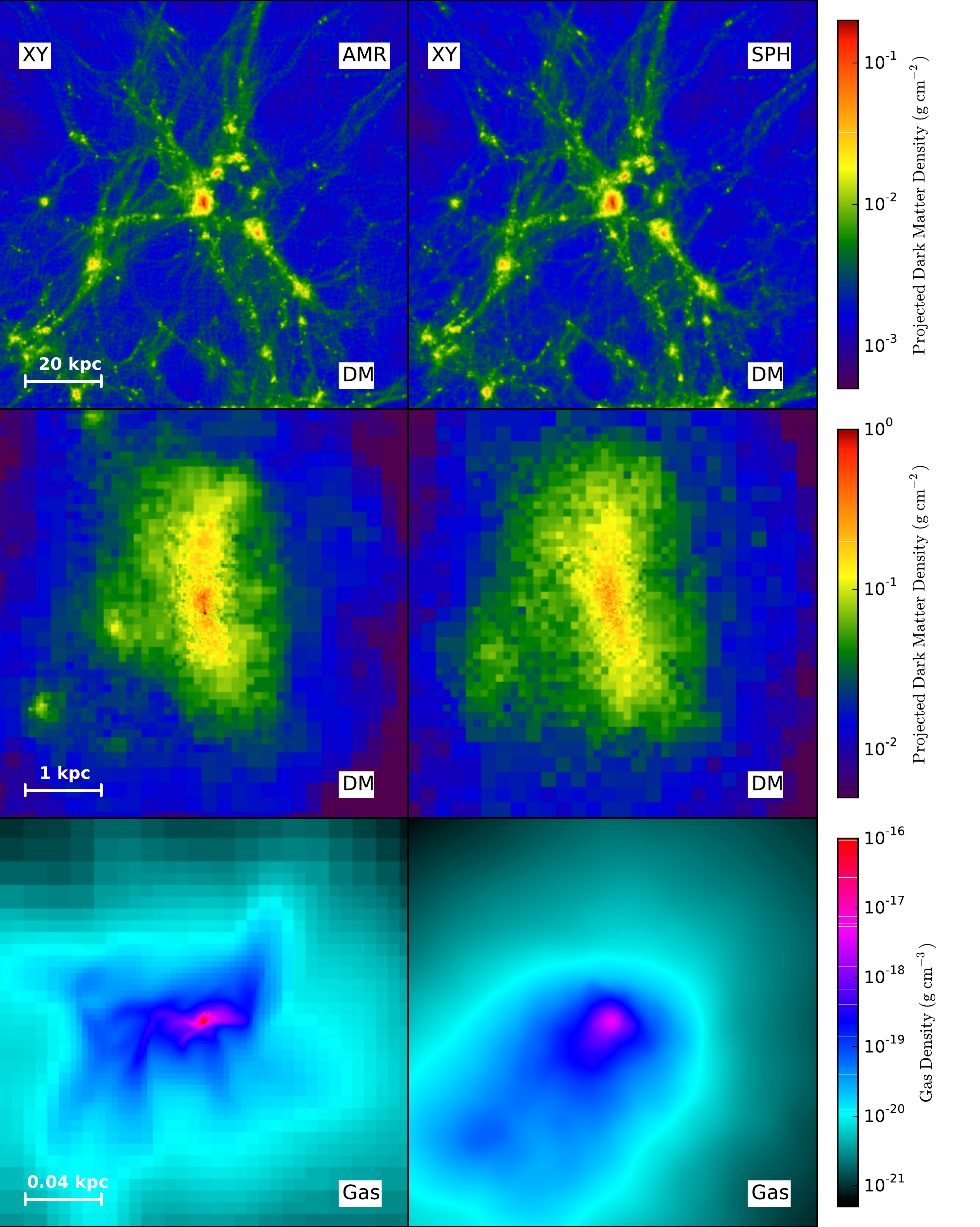}
\caption{Projections of final DM density in $x-y$ plane in cosmological models AMR-C4.2
(left column) and SPH-C4.2 (right column).
The spatial scale of each panel decreases from top to bottom: DM in 100\,kpc
(comoving), DM in 5\,kpc (comoving), and mass-weighted gas volume density in 200\,pc (comoving),
respectively.
The depth of each frame is 10\,kpc (top), and 0.5\,kpc (middle) in comoving coordinates.
The color palette reflects from minimum to maximum projection densities in each panel.
}
  \label{fig:proj_cosmo}
\end{figure*}

Figure\,\ref{fig:proj_cosmo} compares the projection snapshots for DM and volume densities
for the gas of representative cosmological simulations from identical initial conditions
for Enzo (left panels) and GADGET-3 (right panels). The extracted DM halos have $\lambda=0.042$.
The snapshots have been taken at the end of the runs, when the collapse have reached
$R_{\rm fin}\sim 0.01$\,pc.  The top panels show the large scale
distribution of the DM on scales of $100\,h^{-1}$\,kpc (comoving).
Some differences are noticeable in these frames, and typically can be attributed to
individual DM halos. Indeed, we focus on the representative DM halos in the second row
within (approximately) their virial radii, with the box size of $5\,h^{-1}$\,kpc (comoving).
Again, while overall a nice degree of similarity exists between the two runs,
differences are clear as well, especially if one focuses on the substructure on this scale.
The bottom panels show the corresponding gas distribution within these halos,
on scales of $200\,h^{-1}$\,pc (comoving).
While the AMR panel displays the characteristic filamentary structure, the SPH panel shows
a rather centrally-concentrated, elongated gas distribution, although the central density is
the same in both runs. The SPH density distribution displays less small-scale structure
in general. The AGORA collaboration \citep{Kim.etal:14} has also reported difference
in substructure within the similar large-scale structure.

\begin{figure*}
  \centering
\includegraphics[width=0.85\textwidth]{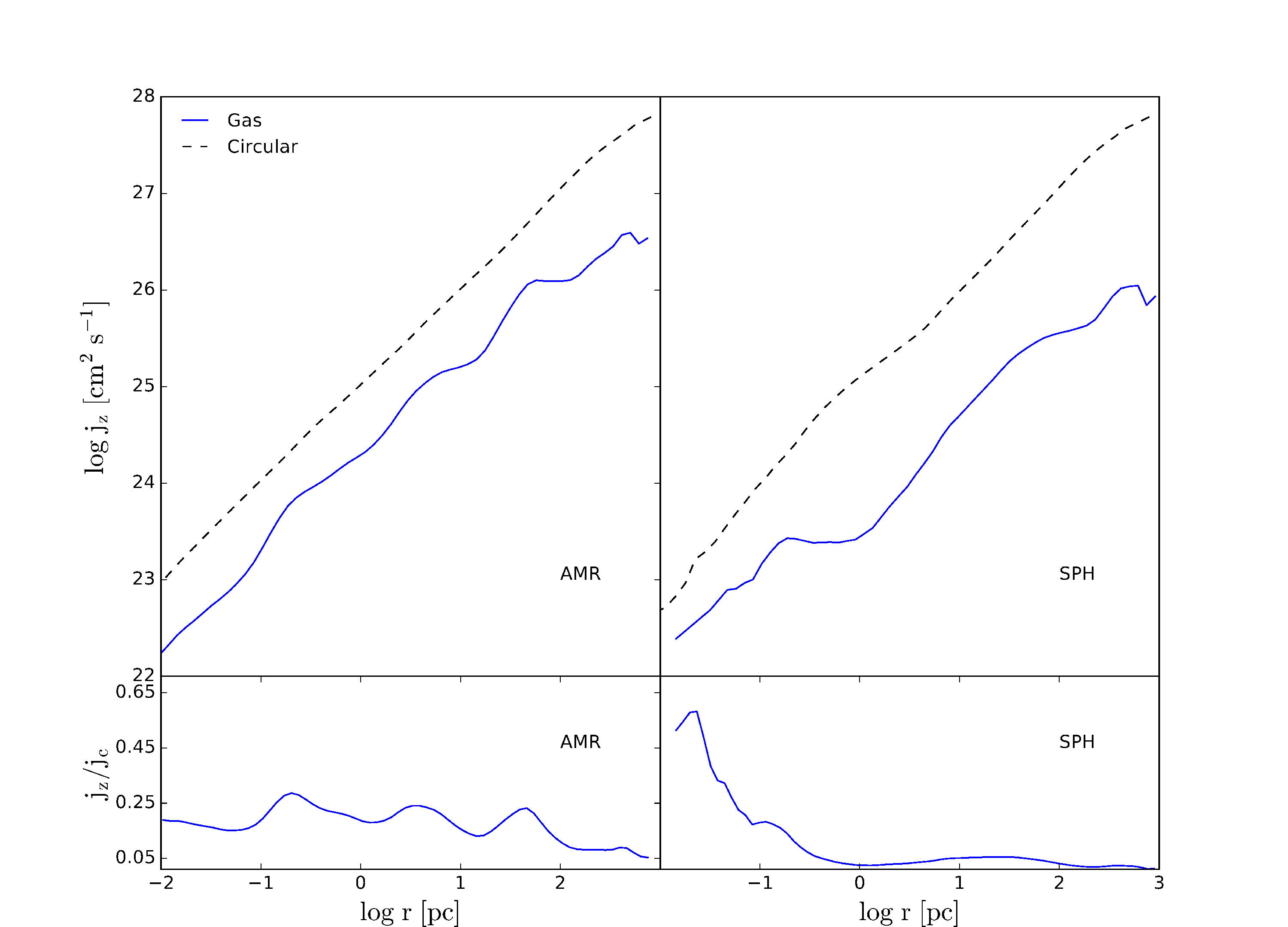}
  \caption{Specific angular momentum profile of gas as function of cylindrical radius
     for Enzo (left panel) and GADGET3 (right panel) in the cosmological runs AMR-C4.2 and SPH-C4.2.
     Circular angular momentum profile is shown for comparison (dashed line).}
  \label{fig:cosmo_jz}
\end{figure*}

Next, we show the radial distribution of the specific angular momentum
profile at the end of the simulations, as well as $j_{\rm z}/j_{\rm c}$ ratio
(Fig.\,\ref{fig:cosmo_jz}). Clearly, both runs display
much less angular momentum support at all radii compared to the isolated models.
On top of this, the SPH run shows even less support than the AMR one,
and proceeds basically in the free-fall fashion until it has reached $r\sim 0.1$\,pc.
The explanation for such a dramatic difference between the rotational support in
isolated and cosmological models is related to the efficiency of angular momentum
extraction from the gas by the background DM. In the former models, the DM halos are
nearly axisymmetric, while in the latter case, they are substantially triaxial.
Resulting gravitational torques from the DM onto the gas will be largely
suppressed in the isolated models, and the gas will have stronger rotational support.

\begin{figure*}
  \centering
 \includegraphics[width=1.0\textwidth]{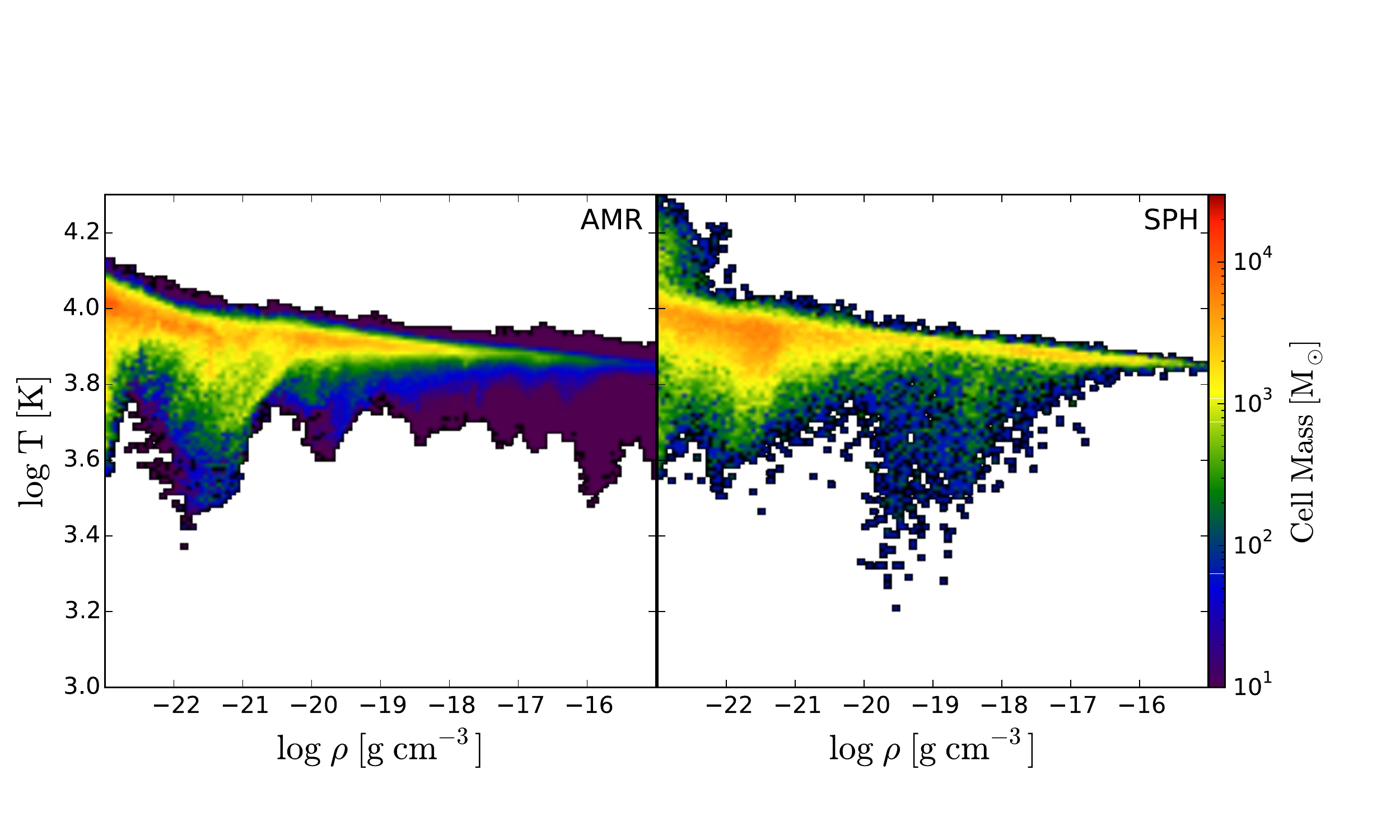}
  \caption{Phase diagram of temperature vs. density in AMR-C4.2 (left panel) and SPH-C4.2 (right panel)
at the end of the cosmological model simulations. The color represents the mass in each pixels. }
  \label{fig:cosmo_density_temperature}
\end{figure*}

We also show the comparison between the $T-\rho$ maps for the above models in
Figure\,\ref{fig:cosmo_density_temperature}.
The overall feature of the phase diagrams are very similar between the two runs,
with the exception of the mass involved at particular $T$ and $\rho$ values.
At high densities, the AMR run shows a somewhat broader distribution in $T$-range
with very small mass contained in each grid-cell, which are completely absent in the SPH run.
This is understandable, as the two codes have a different resolution at these densities.
Although we start from identical initial conditions in Enzo and GADGET-3,
the refinement history differs between the codes in detail.
We have tested how sensitive is the evolution with respect to the refinement condition in Enzo
and discuss it below.

To specify the characteristic times of gravitational collapse in the cosmological models,
we follow the definition from isolated models. The onset of the 1st stage of the  collapse is
inferred from the sudden rise in refinement level in Enzo and from $\eta_{\rm min}(t)$ in GADGET.
The 2nd stage is triggered when the gas-to-DM density ratio exceeds unity in both Enzo and GADGET.

\begin{figure*}
  \centering
\includegraphics[width=1.1\textwidth]{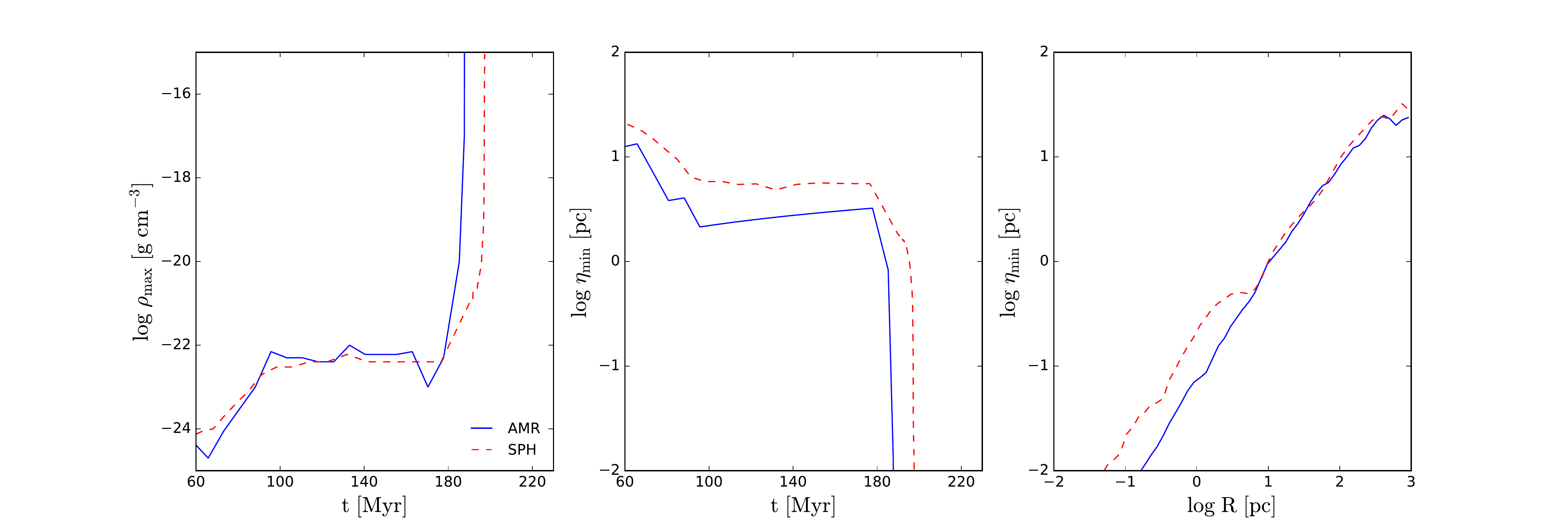}
  \caption{Same as Figure~\ref{fig:compare_res}, but for cosmological models AMR-C4.2 (solid lines)
  and SPH-C4.2 (dashed lines).
Note, that the collapse starts around $t\sim 178$\,Myr for AMR and SPH runs, when $\rho_{\rm
max}$ and $\eta_{\rm min}$ are slightly in favor of the AMR, compared to being identical in the
isolated models. $\eta_{\rm min}(R)$
shows that the innermost resolution is higher in Enzo inside few central pc, while it is
identical in the outer DM halo.
}
\label{fig:cosmo_compare_res}
\end{figure*}

The left panel of Figure\,\ref{fig:cosmo_compare_res} shows the time evolution of
$\rho_{\rm max}$ in cosmological simulations. The time prior to $t\sim 178$\,Myr corresponds
to DM structure formation. It involves the initial expansion and collapse of DM shells,
forming a small DM halo, which we target in the zoom-in simulations. This halo grows via
accretion of cold gas and DM --- a process permeated by merger events. As the halo approaches
the critical mass of $\sim 10^8\,M_\odot$, direct baryonic collapse follows.
The gas collapse starts around $178$\,Myr and proceeds rapidly (but slower than in the isolated
models), being triggered nearly simultaneously in both runs.

The middle frame of Figure\,\ref{fig:cosmo_compare_res} displays the evolution of the minimal
smoothing length, $\eta_{\rm min}$, in the gas. Again, until the onset of the gravitational
collapse, the AMR curve displays a better resolution compared to the SPH one, by a factor of $2-3$,
but both codes are resolving basically the same densities at this time (see \S\,\ref{sec:discuss}
for more details).
This difference appears during the initial collapse and formation of the targeted DM halo.
With the onset of direct collapse at $t\sim 178$\,Myr, this difference in $\eta_{\rm min}$ gets
amplified and the collapse proceeds increasingly faster in the AMR run. This is a reverse situation
with respect to isolated models, and shows up explicitly in the radial profiles of $\eta_{\rm min}$.
We expect that the run with a smaller $\eta_{\rm min}$ will collapse first. We note that the
duration of the baryonic collapse here, $\sim 10-20$\,Myr, is substantially longer than in respective
isolated models. The reason for this difference comes from the collapse inside NFW halos which
have a substantially lower central DM densities than in similar isothermal spheres used by us
for isolated models.

The right frame of Figure\,\ref{fig:cosmo_compare_res} displays the radial profile of
$\eta_{\rm min}(R)$ at the end
of cosmological simulations AMR-C4.2 and SPH-C4.2. Note, the bifurcation in $\eta_{\rm min}$ at around
few pc from the center of the DM halo. Whereas the outer halo region is equally resolved with both runs,
the inner one is better resolved with the AMR. This explains why the 2nd stage of the collapse proceeds
faster with Enzo.

\subsection{Effect of host DM halo spin on isolated models}
\label{sec:res_spin}

As additional and a broad test for model comparison, we measure the collapse time in each model
and contrast the Enzo-2.4 and GADGET-3 runs in isolated models.
Effect of gas rotation on gravitational collapse has been also modeled by \citet{Jappsen.etal:09}
in the context of Pop\,III star formation. However, they assumed a {\it rigid} spherical DM halo
which, therefore, cannot absorb angular momentum and/or produce gravitational torques on the gas.

\begin{figure}
  \centering
\includegraphics[width=0.54\textwidth]{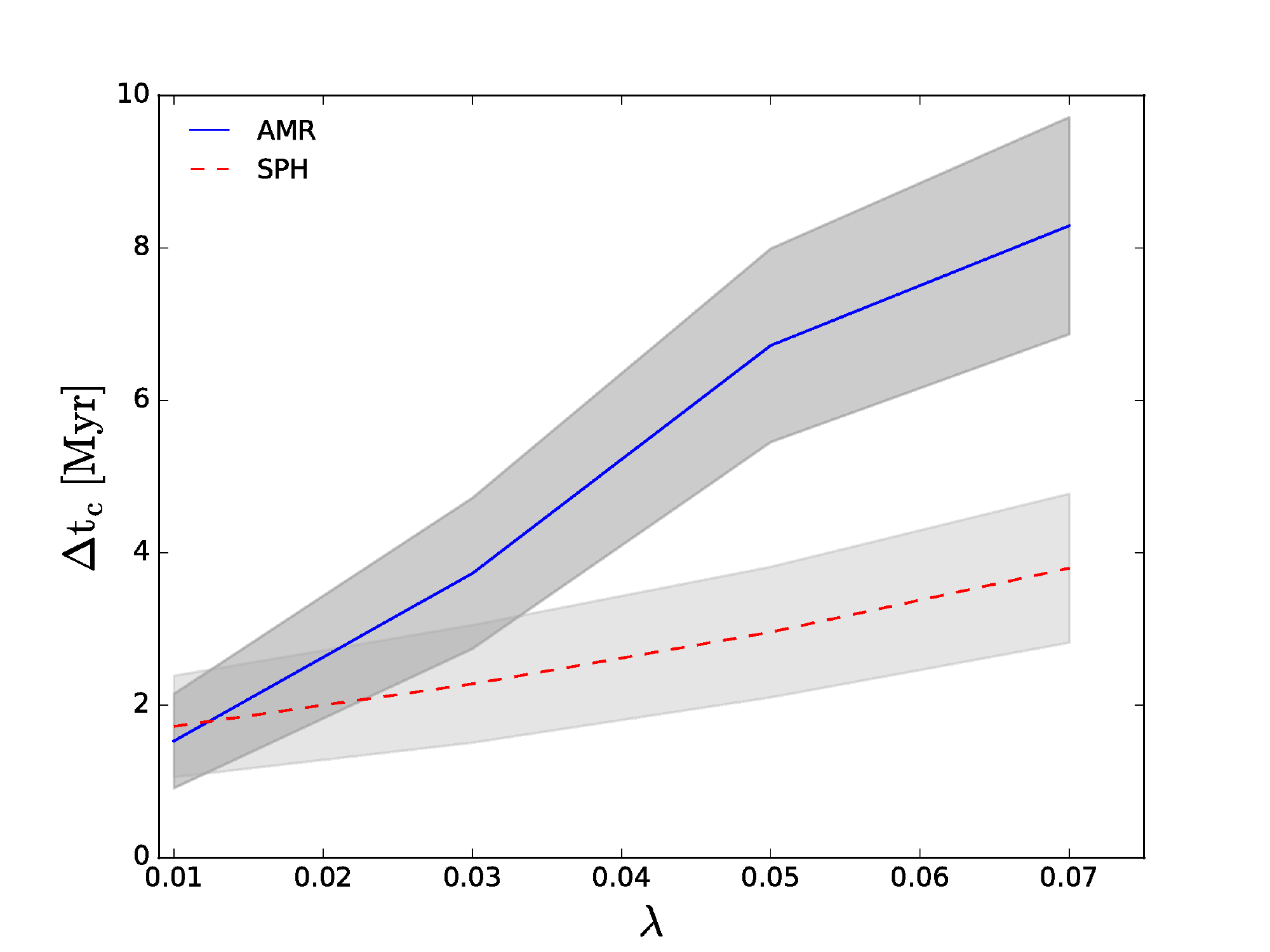}
\caption{Duration of gravitational collapse time for isolated models, $\Delta t_{\rm c}$, in
Enzo-2.4 and GADGET-3 runs as function of the halo spin parameter $\lambda$. The grey
region represents $\pm 1\sigma$ errors using the Poisson statistics.}
  \label{fig:collapsetime_vs_spinparameter}
\end{figure}

To analyze the collapse time for isolated models, we measure the time from the start of the run,
using models with a range in the spin parameter $\lambda$, for AMR and SPH runs.
In other words, we define the collapse time, $\Delta t_{\rm c}$, as the time from $t=0$ till
the collapse
reaches $R_{\rm fin}=10^{-3}$\,pc (\S\,\ref{sec:initial-isolated}).
The collapse time,  $\Delta t_{\rm c}$, is given in Figure\,\ref{fig:collapsetime_vs_spinparameter}.

For isolated models, two trends can be observed.
First, both AMR and SPH runs exhibit a monotonic increase of $\Delta t_{\rm c}$ with $\lambda$,
up to the measured spin of $\lambda\sim 0.07$ (Fig.\,\ref{fig:collapsetime_vs_spinparameter}).
Second, the AMR runs display a longer collapse time than the SPH runs.
As a corollary, the difference between the collapse time
of AMR and SPH runs is increasing monotonically with $\lambda$.

The first trend can be easily explained. With increasing DM halo spin parameter,
the gas must overcome a centrifugal barrier positioned at progressively larger radii,
because the initial conditions for the isolated runs require the gas to
have the same specific angular momentum $j(r)$ distribution as the DM.
The gas evolution is complicated by the intricacies of gravitational collapse:
gas at small radii collapses first and it has lower $j$ than the gas at larger radii.
So the initial circularization determines the position of the accretion shock
which forms at $\sim 1$\,pc from the center, as seen in Figure\,\ref{fig:shock_evolution}.
As the gas with larger $j$ circularizes at larger
$r$ progressively later, the shock moves out, and does it faster for larger $\lambda$.
Hence, the size of the forming disk behind the accretion shock is increasing
with $\lambda$ and with time (e.g., Fig.\,\ref{fig:shock_evolution}).

For the second stage of the collapse to ensue,
the gas must decouple from the background DM potential, i.e., its density must increase above the
DM density at small radii \citep[e.g., \S\,3.1 of][]{Choi.etal:13,Choi.etal:15}. In isolated models
this is approximately the radius of the disk which forms behind the standing accretion shock.
Naturally, a larger accumulation of baryons is required when decoupling happens at larger radii,
explaining the increased $\Delta t_{\rm c}$ for larger $\lambda$. Indeeed, the disk sizes and,
therefore, masses increase with $\lambda$.

Hence, this simple physics explains the relatively modest increase in $\Delta t_{\rm c}$ with $\lambda$.
Remarkably, both Enzo and GADGET-3 runs agree in their $\Delta t_{\rm c}$ for $\lambda=0.01$.
But what about the increasing delay in the collapse time of the AMR runs with $\lambda$?
To understand this, one should look carefully into different resolution pattern of both codes.
While initial conditions are identical in this respect, the SPH gravitational softening is constant
with time, while that of the AMR code are refined as the collapse proceeds.  This by itself
does not have a direct effect on the dynamics, but there are caveats. Probably the most
important issue is the possible effect that refinement has on the development on degree of
turbulence in the collapsing gas.

In Figure\,\ref{fig:compar_smooth}a, we show evolution of $\eta_{\rm min}$ for three representative
$\lambda$ for Enzo and GADGET-3. For $\lambda=0.01$, $\eta_{\rm min}(t)$ displays
little difference between Enzo-2.4 and GADGET-3.
For higher spin, we observe an increasing delay in the 2nd stage of the collapse,
especially for Enzo runs.
Checking the associated increase in the refinement levels, the latter appears to stagnate before the
onset of the 2nd stage of the collapse.
In other words, $\eta_{\rm min}$, which reflects the best spatial resolution at each time,
as mentioned above, ``hesitates" to decrease for a longer time in Enzo compared to
GADGET-3. Therefore, we observe a longer plateau for Enzo in Figure\,\ref{fig:compar_smooth}.

\subsection{Effect of host DM halo spin on cosmological models}
\label{sec:res_cosmo_spin}

In the previous section, we have measured the collapse timescales for isolated models
and contrasted them between Enzo-2.4 and GADGET-3 runs.
Similar exercise in cosmological models is more complicated, because it involves
cosmological evolution of DM halos, which depends on structure formation in the universe
and results in halos having different shapes and other properties.
Hence, while a detailed comparison of the collapse time for cosmological models is outside the
scope of this work, we do compare some numerical aspects of their evolution with various
$\lambda$.

We define the cosmological collapse timescale, $\Delta t_{\rm c}$ by following the evolution
of the baryonic refinement level, the maximal hydrodynamic resolution, $\eta_{\rm min}$,
and the maximal density, $\rho_{\rm max}$ --- all defined as in isolated models. The beginning of
the direct collapse in the cosmological DM halos, $t_0$, is determined when these parameters exhibit
a sharp increase or decrease, correspondingly. For example,  $\eta_{\rm min}$ decreases
sharply at $t\sim 178$\,Myr, which is considered as the onset of the collapse. The end
point of the collapse is taken $t_{\rm fin}$, when it reaches $R_{\rm fin}$, and
$\Delta t_{\rm c} = t_{\rm fin} - t_0$. To detect a sharp increase/decrease in $\rho_{\rm max}$
and $\eta_{\rm min}$, we follow their time derivatives after $\sim 140$\,Myr and test
all the occurrences for the increase/decrease by a factor of 2 or more.

\begin{figure*}
  \centering
\includegraphics[width=0.98\textwidth]{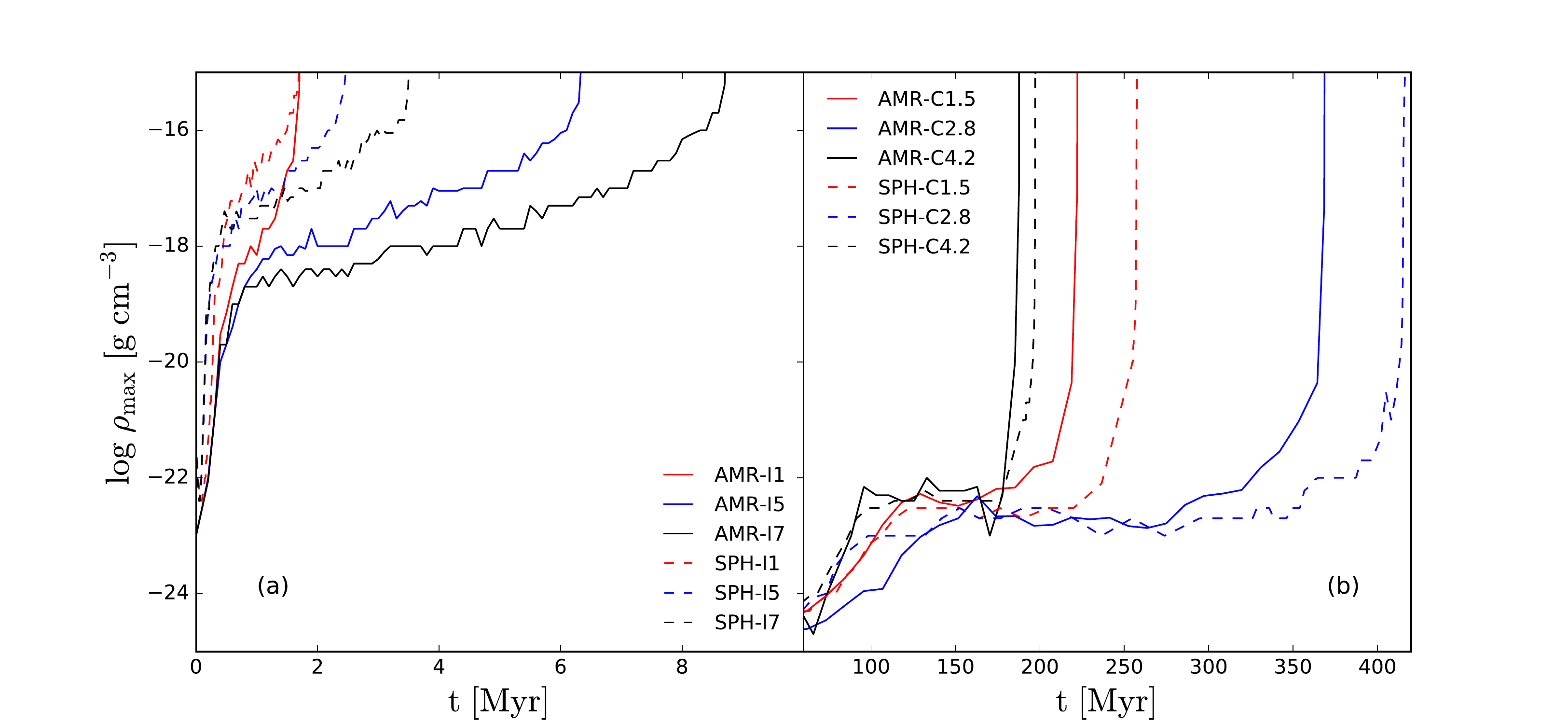}
\caption{Comparison of the evolution of $\rho_{\rm max}(t)$ for {\it (a)} three isolated
models with $\lambda = 0.01, 0.05$ and 0.07 for AMR-I1, -I5 and -I7 (solid lines) and SPH-I1,
-I5 and -I7 (dashed lines) runs. {\it (b)} Same for three cosmological models,
AMR-C1.5, - C2.7 and -C4.2 (solid lines) and SPH-C1.5, -C2.7, and -C4.2 (dashed lines).
The collapse timescales, $\Delta t_{\rm c}$, for these models are listed in
Table\,\ref{tab:collapse_timescales}.}
  \label{fig:compar_smooth}
\end{figure*}

We repeat the $\lambda$-test discussed in the previous section, for cosmological models.
Naively, one expects to observe the same correlation between $\Delta t_{\rm c}$ and $\lambda$
shown in Table\,\ref{tab:collapse_timescales} and in Figure\,\ref{fig:collapsetime_vs_spinparameter}.
However, as exhibited by
Figure\,\ref{fig:compar_smooth}b, this is not the case. For three representative cosmological
models in AMR and SPH, the one-to-one correlation between $\Delta t_{\rm c}$ and
$\lambda$ does not exist here. Why does this happen?

A number of physical reasons in cosmological models would `mess up' this correlation.
First, the host DM halos in isolated models are axisymmetric. This means that collapsing gas
has difficulty to lose angular momentum to DM, and loses it mostly to gas. On the other hand,
the DM halo shapes in cosmological models differ profoundly from axisymmetric shapes. Therefore,
the gas will lose its angular momentum continuously and efficiently to the DM, which
will move the gas away from the centrifugal barrier at each radius, as in fact is shown in
Figure\,\ref{fig:cosmo_jz}.

Secondly, because the direct collapse takes about $10-20$\,Myrs in cosmological models, the
basic properties of DM halos can change during this time. They can undergo minor,
intermediate and major mergers. In other words, the host halos will have enough time to
interact with the substructure --- which does not happen in isolated models.

These and additional factors are expected to affect the direct collapse onset and duration
in cosmological simulations in a kind of unpredicatble way, which we indeed observe in our
simulations.

For isolated and cosmological SPH and AMR models, the free-fall time for the DM halos at
the onset of direct collapse is $t_{\rm ff}\sim 35$\,Myr. In comparison with the actual collapse
timescales given in Table\,\ref{tab:collapse_timescales}, the collapse times are shorter
than the free-fall times. This underlines an important point that direct collapse involves
only the inner part of the gas initially residing in DM halos.

\begin{table}
  \centering
  \caption{Values of $\Delta t_{\rm c}$ for isolated and cosmological models shown in
     Figure\,\ref{fig:compar_smooth}. The definitions of $\Delta t_{\rm c}$ are given in
     \S\S\,\ref{sec:res_spin} and \ref{sec:res_cosmo_spin} for isolated and cosmological models,
     respectively.
  }
  \begin{tabular}[!h]{ccc}
    \toprule
    Models & $\Delta t_{\rm c}$\,(Myr) \\
    \midrule
    AMR-I1 & 1.7  &          \\
    AMR-I5 & 6.3  &          \\
    AMR-I7 & 8.3  &          \\
    SPH-I1 & 1.7  &          \\
    SPH-I5 & 2.3  &          \\
    SPH-I7 & 3.4  &          \\
    \midrule
    AMR-C1.5 & 14.6  &          \\
    AMR-C2.8 & 17.5  &          \\
    AMR-C4.2 & 14.3  &          \\
    SPH-C1.5 & 20.6  &          \\
    SPH-C2.8 & 19.8  &          \\
    SPH-C4.2 & 20.6  &          \\
    \bottomrule
  \end{tabular}
  \label{tab:collapse_timescales}
\end{table}

\section{Discussion \& Conclusions}
\label{sec:discuss}

We have performed comparison simulations of gas collapse in
DM halos in the context of direct collapse scenario to form SMBH seeds using AMR and SPH codes.
This problem involves large dynamic range, from kpc down to AU scales --- such a comparison
has not been attempted before.
Using Enzo-2.4 AMR and GADGET-3 SPH codes, we examined
the systematic differences which buildup when these numerical schemes are invoked.
We made significant efforts to match the initial setup of the two simulations. We addressed the
associated problems using isolated and cosmological models of gravitational collapse in DM halos
with a range of cosmological spin parameter $\lambda$.
To simplify this task, we ignored the effect of molecular hydrogen and UV background radiation,
and assumed an optically-thin atomic cooling, as described in \S\S\,\ref{sec:intro} and \ref{sec:method}.
As a result, the collapse is nearly isothermal in all runs.
Our main results are as follows:

\begin{itemize}
\item
The isolated models generally follow similar evolutionary path and agree with \citet{Choi.etal:13},
but exhibit many subtle differences among themselves. Specifically, the collapse proceeds in two
stages, first reaching the angular momentum barrier, going through an accretion shock, and forming
a disk behind it. In the second stage, which begins after the gas density in the disk region
surpasses that of the DM density, the gas decouples from the background gravitational potential of
the DM and experiences a runaway collapse, dragging some of the DM inward in an adiabatic
compression. We find that, although the AMR and SPH models start with identical initial conditions,
the pace of increase in spatial resolution differs in both codes. This leads to a substantially
earlier collapse in the SPH models for the isolated models.

\item
Cosmological models do not exhibit `standing' accretion shocks, unlike their isolated model
counterparts, and, consequently, do not form rotationally-supported disks. The reason for this is
the continuous loss of angular momentum by the collapsing gas to the background triaxial potential of
the host DM halos, contrary to isolated models which stay largely axisymmetric. Nevertheless, one can
distinguish between first and second stages of the collapse, when the gas-to-DM density ratio
crosses unity, within the central few pc. The cosmological evolution of DM halos and
baryons appears to be similar among the AMR and SPH runs, being only slightly in favor of
the AMR models in baryonic spatial and mass resolutions. But a pronounced trend exists between
the AMR and SPH cosmological models, with the former exhibiting direct collapse at higher redshifts,
and, consequently, the collapse happens in less massive and smaller DM halos.
With the onset of baryonic gravitational collapse within DM halos, the pace of the
resolution change differs among the codes and the resolution increases faster in the AMR runs.
Consequently, the trend in the time-lag is reversed here with respect to the isolated models, and
the AMR models collapse before the SPH models. Thus, a small difference in the pre-collapse resolution
is amplified substantially during the collapse.

\item
The collapse time $\Delta t_{\rm c}$ increases with increasing $\lambda$ in isolated models. Greater
centrifugal forces divert the infalling gas to larger radii, i.e., circularization happens further
away from the rotation axis. This is translated to larger expansion velocities of the accretion
shocks and larger accretion disk sizes. Shock in the AMR runs propagate further out than in the SPH
runs, because the final collapse time is later for the AMR isolated models.

\item For cosmological models, the baryonic collapse time is much longer than in isolated models,
because the average density in the central regions of the NFW halos is lower than that of the
isothermal spheres by a factor of $\sim 100$, which leads to longer collapse time by a factor of
$\sim 10$. Additional factors discussed above affect $\Delta t_{\rm c}$, unlike in isolated models.
\end{itemize}

In this work we have focused on numerical aspects of gravitational collapse to the SMBH seeds,
and so avoid discussing the associated physics, except when unavoidable.
The main reasons for dissimilar performance of the AMR and SPH codes lies in
the mismatch between the pace of increase in spatial and mass resolutions when comparing
performance of the AMR and SPH codes. These differences appear to be profound and not easily
reconcilable.

Both numerical codes under comparison here, AMR Enzo-2.4 and SPH GADGET-3, follow collisionless
cold DM and dissipative fluid component. Each of these components is important in order to understand
evolution of gravitational collapse. The DM plays a crucial role in diluting the gravitational
interactions within the
gas, and, therefore, increasing its Jeans mass. Together with development of virial
supersonic turbulence in the gas, it reduces the fragmentation in the collapsing flow, and allows
the collapse to proceed via large dynamic range.

Most important is the resolution of the gravitational force.
Enzo uses a multi-grid PM method, and therefore, its resolution is governed by
twice the minimum cell size (\S\,\ref{sec:initial-isolated}).
On the other hand, GADGET-3 uses the tree method to compute gravity, and its gravitational resolution
is set by a fixed softening length. Hence, the forces are computed more accurately in GADGET-3
compared to Enzo, particularly at the initial stage of collapse, when Enzo has not refined many levels yet.
In principle, one should be able to make the Enzo gravity resolution comparable to that of GADGET from
the very beginning, by investing significantly larger number of initial mesh allocations, e.g., by nested
initial grid. However, this was not feasible with reasonable computational resources for the present work.

In isolated models, we have been  able to push the initial spatial resolution of
Enzo to $\sim 10$\,pc with three
initial nested grid levels. At the same time, GADGET has had a constant softening length of
$\epsilon_{\rm g}=10^{-4}$\,pc for the SPH particles and $\epsilon_{\rm DM}=0.37$\,pc for the DM, from the
beginning (see Table\,\ref{tab:setup}).  Of course the particles are not as close as these distances
from the start of
the run, but the forces are computed accurately with the tree method down to $\epsilon_{\rm g}$ and
$\epsilon_{\rm DM}$. On the other hand, Enzo gravity was allowed to refine, but it did not refine fast enough
to catch up  with GADGET in the isolated models, judging from the evolution of the hydrodynamical resolution
parameter, $\eta_{\rm min}$.  For periods of time, Enzo refinement level has been
constant, then changed abruptly. Thus, Enzo resolution always lagged behind GADGET in the isolated models.
Due to these differences, Enzo was unable to follow the steep density cusp of the DM isothermal sphere
in the center.

The above situation is similar to the one found in the context of cosmological simulation by
\citet{Oshea.etal:05}, where Enzo and GADGET codes have been compared for cosmological structure formation.
In this case, Enzo had to invest in $\sim 256^3$ root-grid in order to reproduce the same DM
halo mass function as the GADGET run with 128$^3$ particles.  When simply using the root-grid of $128^3$,
Enzo was unable to follow the growth of early density perturbations due to poorer force resolution, and it
resulted in an underestimate of halo substructures and low-mass halos at early times.
This means that Enzo AMR code requires significantly larger computing resources than GADGET SPH code in
order to obtain similar structures on small scales,
particularly in the context of cosmological structure formation.

In the present work, the context and the spatial scale of the problem is very different from that of
\citet{Oshea.etal:05}. Enzo's main advantage is the refinement method which allows it to obtain
a superior hydrodynamic or gravitational force resolution, albeit toward the end of the cosmological
simulation and
in a very small volume. Therefore, during the pre-collapse phase, when high resolution is not required,
the difference in computational resources used by both codes are comparable, within a factor of two,
in the CPU time. On the other hand, during the direct collapse in isolated and cosmological models,
GADGET requires 5 -- 10 times more CPU time than Enzo. Compared to Enzo, GADGET over-resolves in
the initial stage of the collapse, and under-resolves in the final stages, unless SPH particle
splitting method is used or the SPH has a dynamic gravitational softening, as in GIZMO
\citep[e.g.,][]{Heller.Shlosman:94,Hopkins:13}.

The bottom line is that in our simulations of direct baryonic collapse in isolated and cosmological
framework, GADGET requires  greater investment in computational resources, but benefits from
better hydrodynamic and gravitational force resolution in the computational box, while Enzo
achieves a superior resolution in a very small volume.

Note that our cosmological models of direct collapse evolve similarly in the AMR and SPH cases.
The left and middle panels of Figure\,\ref{fig:cosmo_compare_res} show that
direct collapse happens nearly simultaneously in AMR and SPH runs, as exhibited by
the behavior of $\rho_{\rm max}$ and $\eta_{\rm min}$. But these two parameters
differ by a factor of a few until the onset of the direct collapse at $t\sim 178$\,Myr. 
As the collapse develops, we observe that SPH lags behind AMR, unlike in isolated models, where the
SPH models collapse first. The reason for this behavior can be observed in the right panel of
Figure\,\ref{fig:cosmo_compare_res}.
Here the AMR has higher mass and spatial resolution than SPH in the inner few pc at the final phase
of the collapse.

The delay in the collapse time can have a significant impact on the cosmological evolution of direct
collapse, in the presence of an external UV background. The latter can dissociate H$_2$ molecules, which is
considered to be a critical factor for the success of this scenario \citep[e.g.,][]{Omukai:01,
Latif.etal:11,Inayoshi.etal:14,Sugimura.etal:14}.
If the collapse is delayed significantly, it might appear favorably for the direct collapse
scenario because of the higher UVB intensity from nearby star formation, which should be
increasing with time at redshifts $z\sim 20$ to 10.
In our current simulations, Enzo-2.4 and GADGET-3 give mixed results on the cosmological collapse
timescales, as we have discussed above. This happens because the cosmological collapse depends
on realistic properties of DM halos, such as their shapes.
In a more controlled environment of isolated models, GADGET-3
exhibits earlier direct collapse than Enzo-2.4, due to the difference in the early force, mass and
hydrodynamic resolution between the two codes.

In summary, we have performed a comparison between the AMR code Enzo-2.4 and the SPH code GADGET-3 in the
framework of direct collapse to SMBH seeds, using both isolated and fully cosmological models of the
collapse. We have followed the model evolution into a strongly nonlinear regime, when small initial
differences have been substantially amplified. Although the overall model evolution has been found similar
at large, substantial differences also exist at the end of the runs. We find that the main cause for these
differences lies in different evolutionary pace of mass and spatial resolution with time and space.
This results in different abilities of the numerical schemes to resolve hydrodynamics and
gravitational interactions, such as shocks and angular momentum transfer by gravitational torques.

\section*{Acknowledgments}
We thank the Enzo and YT support team for help.
All analysis has been conducted using YT (\citealt{Turk.etal:11}, http://yt-project.org/).
We have used the Grackle chemistry and cooling library
\citep[][https://grackle.readthedocs.org/]{EnzoCollab.etal:14,Kim.etal:14}.
We are grateful to Volker Springel for providing us with the original version of GADGET-3,
and to Jun-Hwan Choi and Long Do Cao for their help in the early phase of this project.
K.N. acknowledges the partial support by JSPS KAKENHI Grant Number 26247022.
I.S. acknowledges partial support from STScI grant AR-12639.01-A.
I.S. and K.N. are grateful to the support from the International Joint Research
Promotion Program at Osaka University.
Support for HST/STScI AR-12639.01-A was provided by NASA through a grant from the STScI, which is operated
by the AURA, Inc., under NASA contract NAS5-26555.
Numerical simulations were in part carried out on XC30 at the Center for Computational Astrophysics, National Astronomical Observatory of Japan,
as well as the VCC  at the Cybermedia Center at Osaka University.

\bibliographystyle{mn}
\bibliography{MyRef}


\end{document}